\documentclass[reprint, aps, prd, floatfix, nofootinbib]{revtex4-1}

\usepackage{amsmath,amssymb,bm,color,float,graphicx,mathrsfs, natbib, appendix}
\usepackage{multirow, Macro}

\def\nG{~{\rm nG}}
\def\Bmpc{\rm B_{1{\rm Mpc}}}

\newcommand{\Eq}[1]{Eq.~\ref{#1}}
\newcommand{\Sec}[1]{Sec.~\ref{#1}}
\newcommand{\Fig}[1]{Fig.~\ref{#1}}
\newcommand{\Tab}[1]{Tab.~\ref{#1}}

\begin{document}
\title{Distinguishing Primordial Magnetic Fields from Inflationary Tensor Perturbations in the Cosmic Microwave Background}
\author{Yilun~Guan}
\email{yilun.guan@utoronto.ca}
\affiliation{Department of Physics and Astronomy, University of Pittsburgh, Pittsburgh, PA, USA 15260}
\affiliation{Dunlap Institute for Astronomy and Astrophysics, University of Toronto, 50 St. George St., Toronto, ON M5S 3H4, Canada}
\author{Arthur~Kosowsky}
\affiliation{Department of Physics and Astronomy, University of Pittsburgh, Pittsburgh, PA, USA 15260}
\date{\today}

\begin{abstract}
  A claimed detection of cosmological tensor perturbations from
  inflation via B-mode polarization of the cosmic microwave background
  requires distinguishing other possible B-mode sources. One such
  potential source of confusion is primordial magnetic fields. For
  sufficiently low-amplitude B-mode signals, the microwave background
  temperature and polarization power spectra from power-law tensor
  perturbations and from a power-law primordial magnetic field are
  indistinguishable. However, we show that such a magnetic field will
  induce a small-scale Faraday rotation which is detectable using
  four-point statistics analogous to gravitational lensing of the
  microwave background. The Faraday rotation signal will distinguish a
  magnetic-field induced B-mode polarization signal from tensor
  perturbations for effective tensor-scalar ratios larger than 0.001,
  detectable in upcoming polarization experiments.
\end{abstract}
\maketitle
% ================== End of Header =========================

\section{Introduction}
One of the primary goals of the next-generation of cosmic microwave
background (CMB) experiments is to detect the primordial B-mode polarization signal
from the tensor perturbations generated by inflation. A detection of
this signal would be compelling evidence of inflation and help
determine the physical mechanism of inflation. While early-universe inflation
generically predicts the production of metric tensor perturbations with a nearly
scale-invariant spectrum via quantum fluctuations
in the gravitational field, the amplitude of the tensor spectrum can vary
greatly between plausible inflation models.

The current best constraint on the tensor-to-scalar ratio is $r<0.056$ at $95\%$
confidence level through a combined analysis of Planck and BICEP2
\citep{Planck2018:X:inflation}. The next generation of large-angle CMB
polarization experiments. including the Simons Observertory
\citep{SO:2019:SciGoal}, BICEP3 \citep{Bicep3:plan}, LiteBIRD
\citep{LiteBIRD:2019}, and CMB-S4 \citep{S4:2016:SciBook} will have
the sensitivity and frequency range to reduce this bound to
$r = 10^{-3}$ or below. However, the tensor perturbations from
inflation are not the only source of B-mode polarization in the CMB.
Foregrounds and lensing, in particular, both are known to contribute
to B-mode polarization. Even regions of the sky with expected 
low galactic foregrounds still have polarized foregrounds which are
substantially larger than current upper limits on any primordial B-mode
polarization component
\citep{BK:V:bmode,PlanckInt:XXX:bicep,BK:X:2018:PGW}. In order to separate
foregrounds from cosmological polarization signals,
the coming generation of large-angle B-mode experiments (BICEP3,
Simons Observatory, LiteBIRD) will measure in many frequency bands, and
test the spatial isotropy and gaussianity of any signal.

We also have known for a long time that the lensing B-mode signal has
a low-$\ell$ contribution whose power spectrum can be mistaken for or
confused with a low-amplitude primordial signal
\citep{Knox:2002:tensor}. B-mode polarization from lensing has
been detected in cross-correlation by SPT \citep{Hanson:2013} and ACT
\citep{ACT:vanEngelen:2015}. We have made great progress at measuring
lensing signals through their non-Gaussian 4-point signature (see,
e.g., \citep{Hu:2001:lensflat}), and now reconstruct maps of the
lensing deflection potential with data from ACT \citep{ACT:Darwish:lensing}, SPT
\citep{Holder:2013:phi}, and Planck
\citep{Planck2018:VIII:lensing}. In principle, this can be done with
very high precision, given clean enough maps with low enough noise
(see, e.g., \citep{Simard:2015:delens,Seljak:2004:Blens}). But in
practice there is a limit to how well low-$\ell$ lensing can be
reconstructed due to having imperfect data with non-zero noise. For
example, although detecting a signal with $r\sim10^{-6}$ is theoretically
achievable in the absence of any systematic errors, sky cuts, and
foregrounds \cite{Seljak:2004:Blens}, realistic forecasts that include
such effects generally predict a much lower sensitivity at the level
of $\sigma(r)\sim 10^{-3}$ \citep{Alonso:2017:bfcast}.

Foregrounds and lensing are the two most important confusion signals
for primordial B-mode polarization, and detailed studies and modeling
of those are well in hand (see \citep{Kamionkowski:2016:review} for a
review). What else could confuse us? Perhaps the next most-likely
signal would be from a primordial magnetic field. Such concern has
previously been brought up in, e.g.,
Refs.~\citep{Brown:2010,Pogosian:2018:PMF}, and discussed in
Ref.~\citep{Renzi_2018}. The extent to which we can distinguish the
two signals, given imperfect data with non-zero noise, motivates this
paper.

Magnetic fields are ubiquitous in the universe today, with typical
strengths of a few microgauss in galaxies and galaxy clusters (see,
e.g.,
Ref.~\citep{Widrow:2002:MF,Durrer:2013,kahniashvili18:magnet_early_univer}
for reviews). Furthermore, evidence from the non-observation of the
inverse Compton cascade $\gm$-rays from the TeV blazars
\citep{2010Sci...328...73N} suggests that magnetic fields are present
in the intergalactic medium, with a lower limit of around
$10^{-7}$\,nG on megaparsec scales. However, the physical origin of
the cosmic magnetic field remains poorly understood. One intriguing
possibility is that cosmic magnetic fields are present before
structure formation and are produced in the very early universe such
as during inflation \citep{PhysRevD.37.2743} or during a phase
transition \citep{Vachaspati:1991nm}. Magnetic fields that are present
before the decoupling of CMB photons are generally known as primordial
magnetic fields.

If present, a primordial magnetic field impacts both the ionization history
of the universe 
and structure formation, leaving imprints on the CMB
and the matter power spectrum \citep{Shaw:2010}. In particular,
primordial magnetic fields source scalar, vector, and tensor metric
perturbations, and influence baryon physics through the Lorentz force.
In addition, primordial magnetic field also introduces a net rotation
of the linear polarization of the CMB photons through an effect known
as Faraday rotation, which leaves an observable frequency-dependent
signal in the CMB polarization pattern
\citep{Kosowsky:1996:FR,Kosowsky_2005}.

The amplitude of the comoving magnetic field $B_0$ present today is
constrained to be no more than a few nG (see, e.g.,
\citep{zucca17,Planck2015:XIX:pmf}). However, it has been previously
shown that a magnetic field with mean amplitude of around 1\,nG and a
power-law power spectrum can generate a CMB B-mode power spectrum
similar to that of an inflationary tensor-mode signal with
tensor-scalar ratio $r\simeq 0.004$ \citep{Renzi_2018}. This is roughly the
limiting tensor amplitude which will be detected by upcoming CMB
experiments. Hence, a lack of knowledge of the primordial magnetic
field may potentially lead us to a wrong conclusion if a B-mode polarization
signal were to be detected by upcoming CMB experiments.

In this work we aim to review and re-evaluate, with particular focus
on the upcoming CMB experiments, the potential degeneracy between a
B-mode signal from a primordial magnetic field model and that from
primordial gravitational waves. 
%(sometimes also referred to as the
%primordial tensor-mode signal in the text). 
In particular, we evaluate
the degeneracy for different tensor-to-scalar ratios $r$, in
the context of experimental configurations that model the
capabilities of upcoming CMB experiments. We also investigate the
extent to which we can break the degeneracy with Faraday rotation from the
magnetic field, at both the power spectrum level and the map level. In
particular, as we shall show in \Sec{sec:ch2:rot}, quadratic
estimation of Faraday rotation at 90\,GHz gives a much more
significant detection of magnetic fields than the power spectrum for a
given map noise and resolution; for a tensor-mode signal at the level
of $r=10^{-3}$, Faraday rotation clearly breaks the power spectrum
degeneracy between tensor perturbations and magnetic fields.

This paper is organized as follows. In \Sec{sec:2}, we review the
basics of the primordial magnetic field. In \Sec{sec:3}, we summarize
the primordial magnetic field contributions to the CMB power
spectrum and evaluate the potential confusion with the tensor-mode
signal from inflation. In \Sec{sec:ch2:faraday}, we briefly review the
physics of Faraday rotation from primordial magnetic field and discuss
to what extent this effect allows us to break the degeneracy between
primordial magnetic field and primordial tensor-mode signals. In
\Sec{sec:ch2:rot}, we summarize the reconstruction of Faraday
rotation through quadratic estimators and then discuss to what extent
it helps us break the degeneracy. Finally, we discuss our results and
conclude in \Sec{sec:ch2:discussion}.

\section{Primordial Magnetic Field} \label{sec:2}
\subsection{Statistics of stochastic magnetic fields}

% ++++++++++++++++++++++++++++++++++++++++++++++++++
\begin{figure*}[t]
  \centering
  \includegraphics[width=0.9\textwidth]{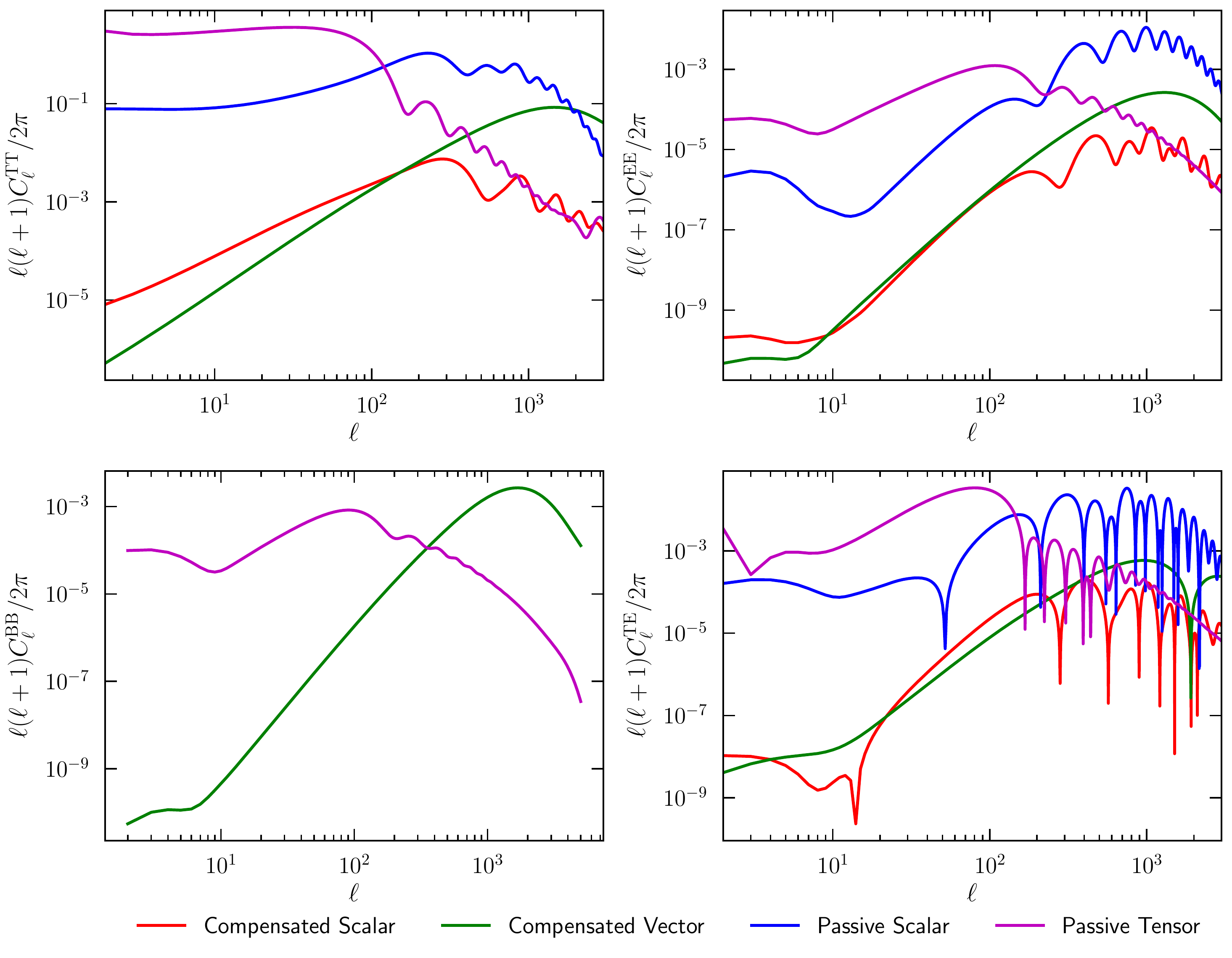}
  \caption{Contributions of different magnetic modes on the CMB power
    spectra (in units of $\mu$K$^2$) from a stochastic background of
    primordial magnetic field with $\Bmpc=1\nG$,
    $\log_{10}\tau_B/\tau_\nu=17$, and $n_B=-2.9$ (nearly scale-invariant)
    generated using \texttt{MagCAMB}.}
  \label{fig:3-1}
\end{figure*}
% ++++++++++++++++++++++++++++++++++++++++++++++++++

We consider a stochastic background of magnetic fields generated prior
to recombination and shall assume that the magnetic field is weak
enough to be treated as a perturbation to the mean energy density
of the universe. As the universe is highly conductive prior to
recombination, so any electric field quickly dissipates. On scales
larger than the horizon, the
magnetic field is effectively ``frozen in'' due to the negligible
magnetic diffusion on cosmological scales. Hence, the conservation
of magnetic flux gives the scaling relation
$B^i(x^j,\tau) = B^i(x^j)/a(\tau)^2$, with $a$ the scale factor,
$\tau$ the conformal time, and $x^j$ the comoving coordinates. We shall
also assume that the stochastic background of magnetic fields follows
the statistics of a Gaussian random field, and the energy density of
magnetic fields, which scales quadratically with the magnetic field
strength ($\propto B^2$), follows chi-square statistics.
In Fourier space,\footnote{In this paper we used the following Fourier convention:\\
  $\tilde{f}(\mathbf{k})=\int d^3 x~
  e^{i\mathbf{k}\cdot\mathbf{x}}f(\mathbf{x})$, and
  $f(\mathbf{x})=\frac{1}{(2\pi)^3} \int d^3 k~
  e^{-i\mathbf{k}\cdot\mathbf{x}}\tilde{f}(\mathbf{k})$.} the statistics
of the magnetic field can be completely described by its 2-point
correlations
% --------------------------------------------------
\begin{equation}
\expect{B_i^*(\mbf{k})B_j(\mbf{k}')} = (2\pi)^3\delta^{(3)}(\mbf{k}-\mbf{k}')[P_{ij} P_B(k) + i\epsilon_{ijl} \hat{k}_l P_H(k)],
\end{equation}
% --------------------------------------------------
where $P_{ij} \equiv \delta_{ij} - \hat{k}_i\hat{k}_j$ is a projection operator
onto the transverse plane to $\hat{k}$ such that $P_{ij}k^j=0$, and
$\epsilon_{ijl}$ is the total anti-symmetric tensor. Here $P_H$ and
$P_B$ refer to the helical and non-helical part of the magnetic field
power spectrum, respectively. For the interests of simplicity we shall
assume that the helical magnetic field component vanishes, though we
should note that helical magnetic field is predicted by some proposed
magnetogenesis scenarios (see, e.g.,
\citep{Semikoz:2005,Campanelli:2005}).

We assume that the power spectrum of magnetic field follows a power
law with a cut-off scale $k_D$, given by
% --------------------------------------------------
\begin{equation}
P_B = A_B k^{n_B},\:\:\:k\le k_D,
\end{equation}
% --------------------------------------------------
which vanishes for $k>k_D$. The dissipation scale $k_D$ reflects the
suppression of magnetic field due to radiation viscosity on small
scales. $A_B$ and $n_B$ denote the amplitude and spectral index of
magnetic field power spectrum, respectively, both of which depend on
the specific magnetogensis scenerio. In particular, an
inflationary magnetogenesis model prefers a scale-invariant spectrum
with a spectral index $n_B \approx 3$, while a causally-generated magnetic
field in the post-inflationary epoch prefers a spectrum with
$n_B\ge 2$ \citep{Pogosian:2018:PMF}.

The assumption that the magnetic field is ``frozen-in'' and follows a
power law with a cut-off scale $k_D$ is only an
approximation. Magnetohydrodynamic simulations (e.g.,
\citep{Brandenburg:1996,Banerjee:2003,Kahniashvili:2016}) have shown
that magnetic fields tend to source turbulence on scales smaller than
the horizon. However, such an effect is expected to have negligible
impact on the results in this paper as it affects mostly the
small-scale magnetic fields, whereas, as we shall discuss in
\Sec{sec:3}, only the large-scale magnetic modes are degenerate with a
primordial tensor mode signal. Thus we neglect subhorizon plasma
dynamics.

In addition, following a common convention in the literature, we smooth the
magnetic field with a Gaussian kernel
$f_\lambda(x) = N\exp{(-x^2/ 2\lambda^2)}$ on a comoving scale $\lambda$.
The magnetic field fluctuation on the comoving scale 
$\lambda$ can then be characterized by 
% --------------------------------------------------
\begin{equation}
B_\lambda^2\equiv \expect{\mbf{B}_\lambda(\mbf{x})\cdot\mbf{B}_\lambda (\mbf{x})} = \frac{1}{\pi^2}\int_0^\infty dk~k^2 e^{-k^2\lambda^2}P_B(k),
\end{equation}
% --------------------------------------------------
which relates to the power spectrum amplitude $A_B$ by
% --------------------------------------------------
\begin{equation}
A_B = \frac{(2\pi)^{n_B+5}B_\lambda^2}{2\Gamma(\frac{n_B+3}{2})k_\lambda^{n_B+3}}.
\end{equation}
% --------------------------------------------------
The damping scale $k_D$ can also be approximated as \citep{Mack_2002},
\begin{equation}
  \begin{split}
    k_D &= (5.5\times10^4)^{\frac{1}{n_B+5}}\left(\frac{B_\lambda}{1\,{\rm nG}}\right)^{-\frac{2}{n_B+5}}\left(\frac{2\pi}{\lambda/{\rm Mpc}}\right)^{\frac{n_B+3}{n_B+5}}\\
    &\times h^{\frac{1}{n_B+5}}\left(\frac{\Omega_bh^2}{0.022}\right)^{\frac{1}{n_B+5}}\vert_{\lambda = 1{\rm Mpc}} ~{\rm Mpc}^{-1},
  \end{split}
\end{equation}
with $h$ the reduced Hubble parameter defined as $h\equiv H_0/100$\,km\,s$^{-1}$\,Mpc$^{-1}$.

% ==================================================
\subsection{Magnetic Perturbations}
% ==================================================
Consider a particular realization of a stochastic magnetic
field, with the magnitude of the field at $x$ and conformal time
$\tau$ given by $B^i({\mathbf x},\tau)$. Its energy-momentum tensor can be
written as
% --------------------------------------------------
\begin{equation}
\begin{split}
  T^0{}_0 &= -\frac{1}{8\pi a^4}B^2(\mathbf{x}), \\
  T^0{}_i &= T^i{}_0= 0, \\
  T^i{}_j &= \frac{1}{4\pi a^4}\left[\hf B^2(\mathbf{x})\delta^i{}_j-B^i(\mathbf{x})B_j(\mathbf{x})\right],
\end{split}
\end{equation}
% --------------------------------------------------
where we have used the ``freeze-in'' condition
$B^i(\mathbf{x},\tau) = B^i(\mathbf{x})/a(\tau)^2$. Since
$\rho_B\equiv T^0{}_0 \propto a^{-4}$ scales the same way as photon energy density,
one can reparametrize the magnetic field perturbation relative to the
photon density $\rho_\gamma$ and pressure $p_\gamma$ as \citep{Shaw:2010}
\begin{equation}
\begin{split}
  T^0{}_0 &= -\rho_\gamma\Delta_B,\\
  T^i{}_j &= p_\gamma(\Delta_B\delta^i_j+\Pi^i{}_{Bj}),\\
\end{split} 
\end{equation}
where $\Delta_B$ denotes the scalar perturbations sourced by magnetic
fields relative to the radiation energy density, and $\Pi^i{}_{Bj}$
denotes the anisotropic stress from magnetic fields which can be
further decomposed into scalar, vector, and tensor type perturbations.

Initial conditions of magnetically-induced
perturbation modes can be decomposed into three types: (1) compensated
\citep{Giovannini:2004,Finelli:2008}, (2) passive
\citep{Lewis:2004ef,Shaw:2010}, and (3) inflationary \citep{Bonvin:2013}.
In particular, compensated magnetic modes arise when
the magnetic contributions to the metric perturbations are compensated
by fluid modes to the leading order on super-horizon scales. It
includes the contributions from magnetic field after neutrino
decoupling, and is finite in the $\tau\rightarrow0$ limit. The passive magnetic
modes, on the other hand, account for the magnetic contribution prior to
neutrino decoupling. In this period, the universe is dominated by a
tightly-coupled radiative fluid which prevents any anisotropic stress
from developing. Without neutrino free-streaming, the magnetic field acts
as the only source of anisotropic stress, leading to a logarithmically
growing mode \citep{Lewis:2004ef}. This mode
survives neutrino decoupling as a constant offset on the amplitude of
the non-magnetic mode. Inflationary magnetic modes, as another type of
initial condition, depend on the specific generation mechanism
\citep{Bonvin:2013}, and is therefore not considered in this paper in
order to maintain generality of our results to different magnetic
field models.

From this physical picture it is apparent that the size of the
perturbations from magnetic fields depends on the epoch of their
generation relative to the epoch of neutrino decoupling, as can be
parametrized by $\log_{10}(\tau_\nu/\tau_B)$, with $\tau_\nu$ the neutrino
decoupling time and $\tau_B$ the magnetic field generation time. Though
the exact number for this quantity remains unknown and can be
model-dependent, we shall assume $\log_{10}(\tau_\nu/\tau_B)=17$ for
simplicity, following Ref.~\citep{Planck2015:XIX:pmf}. This is,
nevertheless, without a loss of generality, as $\tau_\nu/\tau_B$ can be
degenerate with the amplitude of the perturbations (e.g., $B_\lambda$ or
$A_B$) \citep{Shaw:2010}.
In addition, the magnetic field also introduces a Lorentz force acting on
the baryons in the primordial plasma. It effectively augments the
pressure perturbations of the baryon-photon fluid which prevent
photons and baryons from falling into their gravitational wells. This
effect is analogous to a change in baryon energy density which affects
the sound speed of the baryon-photon fluid and changes their acoustic
oscillations \citep{Adams:1996,Kahniashvili:2006,Kunze:2011}.

\section{Impacts on CMB power spectra}
\label{sec:3}

A primordial magnetic field influences CMB anisotropies through both its
metric perturbations and the Lorentz force, and generates
perturbations of scalar, vector, and tensor types. 
We make use of the
publicly available code
\texttt{MagCAMB}\footnote{https://github.com/alexzucca90/MagCAMB}
\citep{zucca17} which extends the Boltzmann code \textsc{CAMB}
\citep{Lewis_2000} to include the effects of a primordial magnetic field
discussed in \Sec{sec:2}. In \Fig{fig:3-1} we show an example
set of CMB power spectra that are sourced by a stochastic 
primordial magnetic field with $B_{1{\rm Mpc}} = 1$\,nG and a
nearly scale-invariant spectrum ($n_B=-2.9$). 
Contributions from different magnetic modes are plotted in
different colors, from which one observes that the passive tensor-mode
signal in $\ClBB$ has significant power at $\ell\lesssim100$ resembling that of
an inflationary tensor-mode signal and hence may pose a possible
source of confusion. On the other hand, the compensated vector-mode
contribution dominates at $\ell \gtrsim 1000$ in both $\ClTT$ and
$\ClBB$ which is not degenerate with the inflationary tensor-mode
signal. Hence, this vector-mode perturbation from primordial magnetic
field gives us a potential handle to break the degeneracy.

To evaluate the extent of the confusion for upcoming CMB
experiments, we simulate different sets of CMB power spectra using
\textsc{CAMB} with the standard \lcdm~model and the Planck best-fit
cosmological parameters as our fiducial model \cite{Planck2018:VI:CP},
while varying the tensor-to-scalar ratio $r$ to reflect different
science targets, with the spectral index $n_T$ fixed by the slow-roll
inflation consistency relation $n_T=-r/8$. We consider several
toy-model full-sky microwave background experiments specified by
angular resolution and map sensitivity. 
In addition, we simulate the observed power
spectra for each experiment with an idealized noise model given by
% --------------------------------------------------
\begin{equation} \label{eq:ch2:nl}
N_\ell = w^{-1} \exp \left(\ell(\ell+1) \theta^{2} / 8 \ln 2\right),
\end{equation}
% --------------------------------------------------
where $w^{-1/2} \equiv \sqrt{4\pi \sigma_{\rm pix}^2 / N_{\rm pix}}$ denotes the
expected noise level of an experiment, with $\sigma_{\rm pix}$ the
per-pixel noise level, $N_{\rm pix}$ the total number of pixels, and
$\theta$ the full-width-half-minimum (FWHM) size of a Gaussian telescope beam.
We also assume that the
polarization and temperature noise are related simply by
$(\sigma^{\rm P}_{\rm pix})^2=2(\sigma^{\rm T}_{\rm pix})^2$.

In \Tab{tab:3-1} we list the toy-model experiments considered in this
work. In particular, Expt~A and B approximate the capabilities of the
Simons Observatory Large Aperture Telescope (SO LAT) and Small
Aperture Telescope (SO SAT), respectively. Expt~C1 represents a
combined constraint with both of these experiments. Expt~C2 represents
the capability of the anticipated CMB-S4 experiment, while C3 is the
limit of a noiseless CMB map so that the power spectrum uncertainty is due 
entirely to cosmic variance. 

% ++++++++++++++++++++++++++++++++++++++++++++++++++
\begin{table}[tbp]
  \centering
  \begin{tabular}{c|c|c|c|c|c}\hline\hline
    Name & Beam [arcmin] & Noise [$\mu$K\,arcmin] & $\ell_{\rm min}$ & $\ell_{\rm max}$ & $f_{\rm sky}$ \\\hline
    A & 17 & 2 & 30 & 1000 & 0.1\\\hline
    B & 1.4 & 6 & 30 & 3000 & 0.4\\\hline
    \multirow{2}{*}{C1} & 17 & 2 & 30 & 1000 & 0.1\\\cline{2-6}
                           & 1.4 & 6 & 30 & 3000 & 0.4\\\hline
    \multirow{2}{*}{C2} & 17 & 1 & 30 & 1000 & 0.1\\\cline{2-6}
                           & 1.4 & 2 & 30 & 3000 & 0.4\\\hline
    \multirow{2}{*}{C3} & 17 & 0 & 30 & 1000 & 0.1\\\cline{2-6}
                           & 1.4 & 0 & 30 & 3000 & 0.4\\\hline\hline
  \end{tabular}
  \caption{Different sets of experimental parameters considered in this
    paper. Expt~A represents a ground-based small-aperture telescope,
    while Expt~B represents a ground-based large-aperture telescope.
    C1, C2, and C3 represent a combination of Expt~A and B at various
    noise levels.}
  \label{tab:3-1}
\end{table}
% ++++++++++++++++++++++++++++++++++++++++++++++++++

We compute Markov Chain Monte Carlo (MCMC) model fitting to find the
best-fit cosmologies for two competing models: (1) a model with a
non-zero tensor-to-scalar ratio $r$ but no primordial magnetic field
contribution (\lcdm+r hereafter); (2) a model with $r=0$ but non-zero
primordial magnetic field contribution (\lcdm+PMF hereafter). The
Markov Chain varies the standard cosmological parameters, plus either
the tensor-scalar ratio or the primordial magnetic field amplitude and
power spectrum index (see Appendix~\ref{app:post} for more details on
the MCMC model fitting and example results from the Markov
Chains). The log-likelihood for a given model is taken as
\citep{Hamimeche:2008:like}
% --------------------------------------------------
\begin{equation}
\begin{split}
  -2 \ln \mathcal{L}\left(\left\{\bhCl\right\} |\{\bCl\}\right)=
  &\sum_{l}(2l+1)\bigg\{{\rm Tr}\left[\bhCl \bCl^{-1}\right] \\
  &- \ln \left|\bhCl \bCl^{-1}\right|-3\bigg\},
\end{split}
\end{equation}
% --------------------------------------------------
where $\bCl$ contains the theory power spectra given by
% --------------------------------------------------
\begin{equation}
  \bCl \equiv \begin{pmatrix}
  \ClTT & \ClTE &     0 \\
  \ClTE & \ClEE &     0 \\
      0 &     0 & \ClBB \\
  \end{pmatrix},
\end{equation}
% --------------------------------------------------
and $\bhCl$ contains the observed power spectra given by
% --------------------------------------------------
\begin{equation}
  \bhCl \equiv \frac{1}{2\ell+1} \sum_{m} {\bm a}_{\ell m} {\bm a}_{\ell m}^\dagger,
\end{equation}
% --------------------------------------------------
with ${\bm a}_{\ell m} \equiv (\almT ~~ \almE ~~ \almB)^T$. Note that
the full set of power spectra, $\ClTT$, $\ClEE$, $\ClBB$, and $\ClTE$
are used in the model-fitting. 

Specifically, the simulated power spectrum is generated with the
\lcdm+r model, which we then fit with a \lcdm+PMF model to find
degenerate magnetic field models in terms of CMB power
spectra. Although in theory the expected power spectra from the two
competing models are not completely degenerate due to, for instance,
the vector-mode signal from the primordial magnetic field, in practice
the difference may not be detectable at a given experimental noise
level, especially when $\Bmpc\lesssim 1$\,nG.  By computing the
$\Delta \chi^2$ between the two best-fit models, we evaluate the
extent of the degeneracy between the \lcdm+r model and the \lcdm+PMF
model at various $r$ targets and experiment sensitivities as listed in
\Tab{tab:3-1}.

% ===================================================
\subsection{Fiducial cosmology with $r=0.01$} \label{sec:3.1}
% ===================================================

% ++++++++++++++++++++++++++++++++++++++++++++++++++
\begin{figure}[t]
  \centering
  \includegraphics[width=0.45\textwidth]{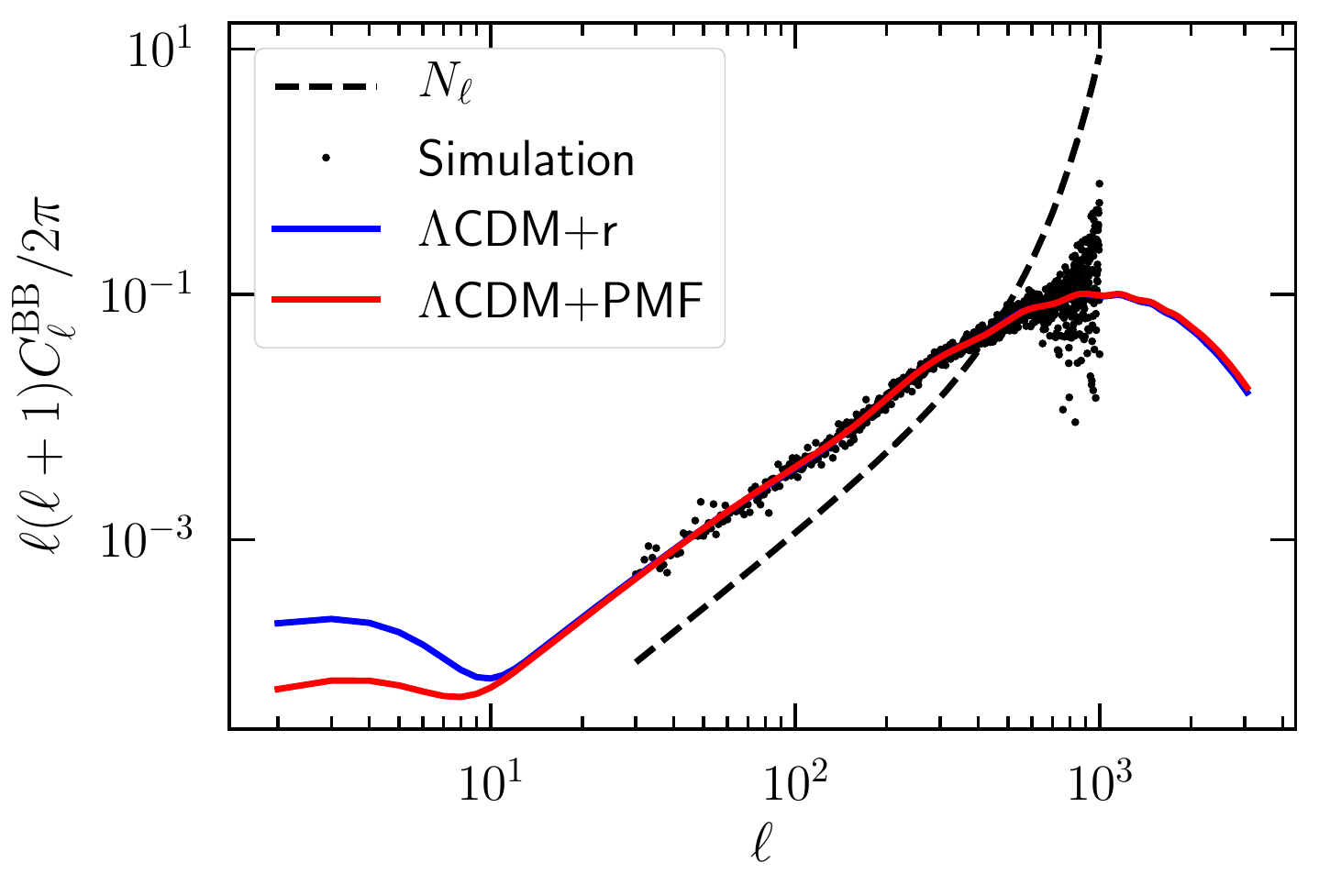}
  \caption{Best-fit $C_\ell^{\rm BB}$ power spectra (in units of $\mu$K$^2$) for
    the \lcdm+r (blue) and \lcdm+PMF (red) models, to an underlying tensor cosmology
    with $r=0.01$ and map noise Expt~A in \Tab{tab:3-1}. The black dots
    represent the simulated data after removing noise model UNCLEAR, and the
    black dashed line represents the noise model.}
  \label{fig:bfps-sat}
\end{figure}
% ++++++++++++++++++++++++++++++++++++++++++++++++++

% ++++++++++++++++++++++++++++++++++++++++++++++++++
% difference of best-fit power spectra
\begin{figure*}[t]
\centering
\includegraphics[width=0.9\textwidth]{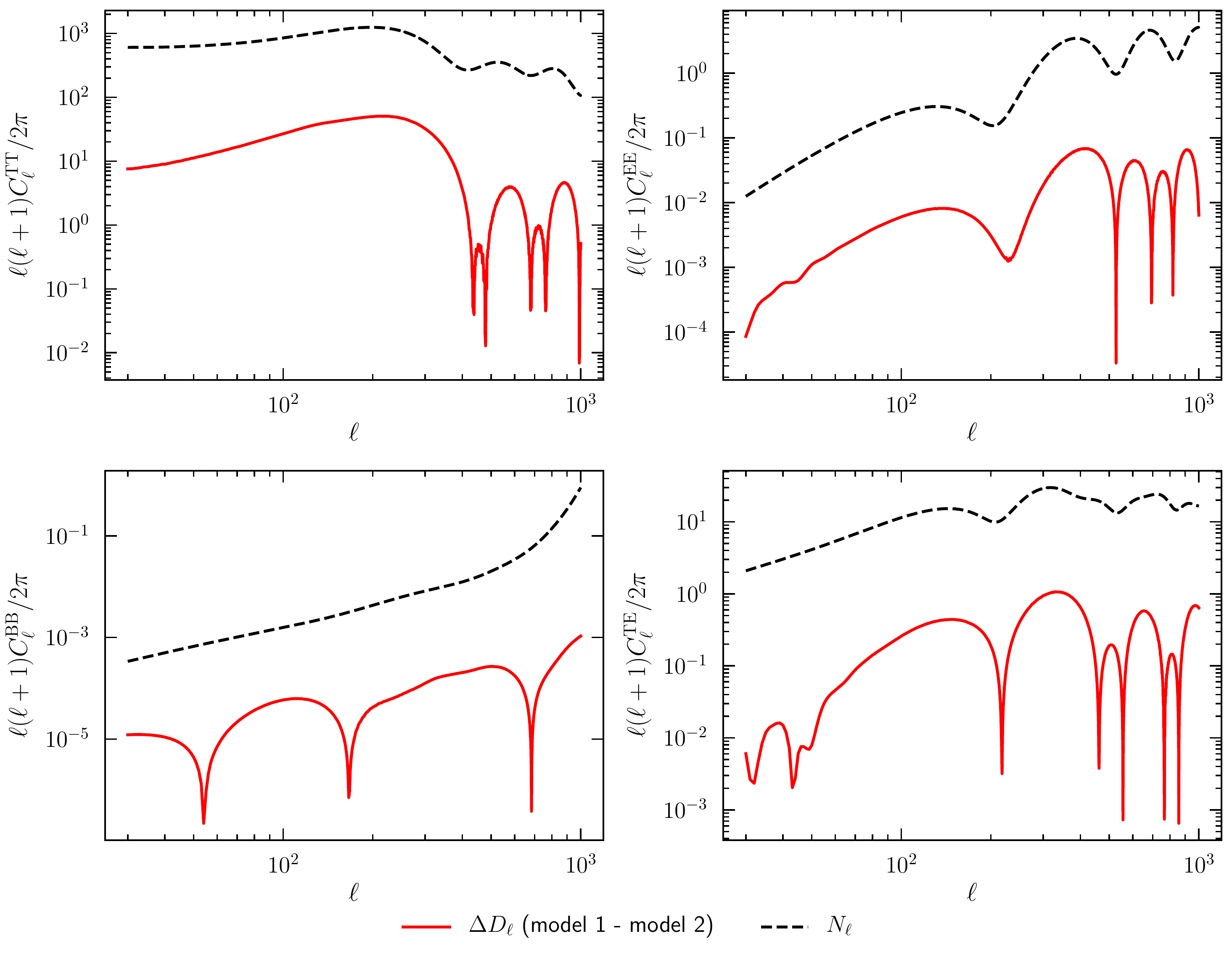}
\caption{Difference of the two best-fit CMB power spectra in Fig.~\ref{fig:bfps-sat}
  is shown by the red solid
  line. The black dashed
  line shows the analytic covariance of the simulated power spectrum.}
\label{fig:diff-sat}
\end{figure*}
% ++++++++++++++++++++++++++++++++++++++++++++++++++

We first consider a target of $r=0.01$ which is one of the primary
goals of the upcoming CMB experiments such as the Simons Observatory
(SO) \cite{SO:2019:SciGoal}. In particular, SO will have two separate
instruments for measuring different angular scales of the CMB power
spectrum: a large-aperture telescope (LAT) which mainly focuses on
small-angle CMB anisotropies, and a small-aperture telescope (SAT)
which mainly focuses on the large-angle CMB anisotropies. As the
tensor-mode signal from inflation is expected to show up predominantly
in the large angular scales, it is the main target of the SO SAT
experiment.

Suppose that we live in a universe well described by a \lcdm+r model
with $r=0.01$, and we measure the CMB power spectrum with an SO
SAT-like experiment, specified as Expt~A in \Tab{tab:3-1}. 
We simulate the observed CMB power
spectra for Expt~A between angular scales of $\ell_{\rm min}=30$ and
$\ell_{\rm max}=3000$, with a sky fraction of $f_{\rm sky}=0.3$, to
account for the effect of partial sky coverage from a ground-based
experiment.

We then fit the simulated data with both the \lcdm+r and the \lcdm+PMF
models. The resulting B-mode polariation power spectra $\ClBB$ for
the two best-fit models are shown in \Fig{fig:bfps-sat}, compared
to the simulated data. It shows that the two competing models can be
highly degenerate over the angular scales probed by the simulated
experiment (Expt~A; $30 \lesssim \ell \lesssim 3000$), with a difference much smaller
than the variance of the observed data. To be more specific, one can
model the variance of the observed data as \cite{Kamionkowski_1997}
% --------------------------------------------------
\begin{equation}
\label{eq:sigma2}
\sigma^{2}\left(C_{\ell}\right)=\frac{2}{(2 \ell+1) f_{\mathrm{sky}}}\left(C_{\ell}+N_{\ell}\right)^{2},
\end{equation}
% --------------------------------------------------
and compare it to the difference between the two sets of best-fit
power spectra, as shown in \Fig{fig:diff-sat}. 
The difference in the best-fit power spectra is around $2$ orders of
magnitude below the expected variance of the observed power spectrum,
indicating that breaking the degeneracy between the two models is
impossible without additional information. The corresponding difference
in $\chi^2$ between these two best-fit models is $\Delta \chi^2 \simeq 0.1$

The degeneracy between the two models is not too surprising
because on large angular scales ($\ell \lesssim 100$) the passive
tensor mode dominates over the other contributions from the primordial
magnetic field, and the passive tensor mode is mathematically
equivalent to the inflationary tensor-mode signal; the degeneracy
is unavoidable if one observes only at the large angular scales. On
the other hand, one does see noticeable difference between the two
models at $\ell \lesssim 10$, indicating that the two models are not completely
degenerate on all angular scales. This is expected because in the
small angular scales ($\ell \gtrsim 1000$), the compensated vector mode signal
from a primordial magnetic field starts to dominate over the other magnetic modes in
the $\ClBB$ power spectrum. This difference in the small scales 
gets minimized by the best-fit model, leading to the difference seen at
$\ell \lesssim 10$. This also implies that the small-scale CMB anisotropies
contain crucial information that helps break the degeneracy
between $r=0.01$ in tensor perturbations and a primordial magnetic field. 

% ++++++++++++++++++++++++++++++++++++++++++++++++++
\begin{figure}[tbp]
  \centering
  \includegraphics[width=0.45\textwidth]{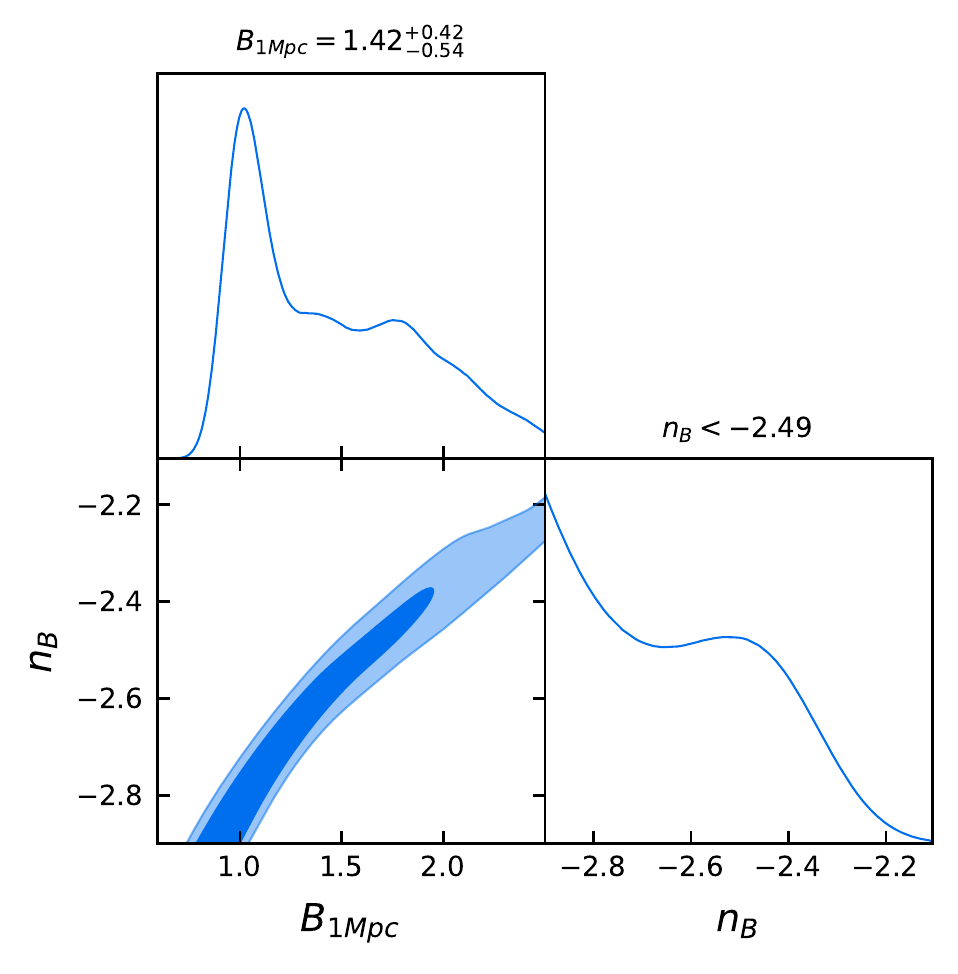}
  \caption{Joint posterior distributions for the \lcdm+PMF model parameters after
    fitting the simulated data (generated with a \lcdm+r model with
    $r=0.01$) to a \lcdm+PMF model.}
  \label{fig:post4_sat}
\end{figure}
% ++++++++++++++++++++++++++++++++++++++++++++++++++

In \Fig{fig:post4_sat} we show the posterior distributions of the
magnetic field parameters ($\Bmpc$ and $n_B$) from the \lcdm+PMF
model fitting an inflationary tensor perturbation at $r=0.01$. 
Specifically, we obtain a best-fit primordial magnetic field
model with $\Bmpc=1.42^{+0.42}_{-0.54}$\,nG at 68\% confidence level,
on par with the observational constraints set by Planck in 2015
\cite{Planck2015:XIX:pmf}. We also note that a nearly scale-invariant
spectrum, with a spectral index of $n_B < -2.49$, is preferred by the
simulated data, which we find a generic feature of the 
\lcdm+PMF models degenerate to \lcdm+r. An apparent degeneracy between
the amplitude of the magnetic field $\Bmpc$ and the magnetic spectral
index $n_B$ can also be seen. This is because as $n_B$ increases, the
power spectrum of primordial magnetic field tilts toward the smaller
scales, leading to less power in the large scale modes which Expt~A
(or an SO SAT-like experiment) is sensitive to, and thus the loss of
power gets compensated by a stronger magnetic field.

Now suppose that one obtains additional observations from a
large-aperture telescope like the SO LAT, specified as Expt~B in
\Tab{tab:3-1}, which strongly constrains the small-scale CMB
anisotropies. One can then combine its constraining power
with Expt~A to jointly constrain the primordial magnetic field on both
small and large angular scales. For simplicity, we simulate the
observed power spectra of the combined constraint by simulating two
separate experiments with the same underlying CMB realization and
combining them trivially by using the experiment that gives the lowest
variance at each $\ell$ to avoid mode double counting.

\begin{figure}[t]
  \centering
    \includegraphics[width=0.4\textwidth]{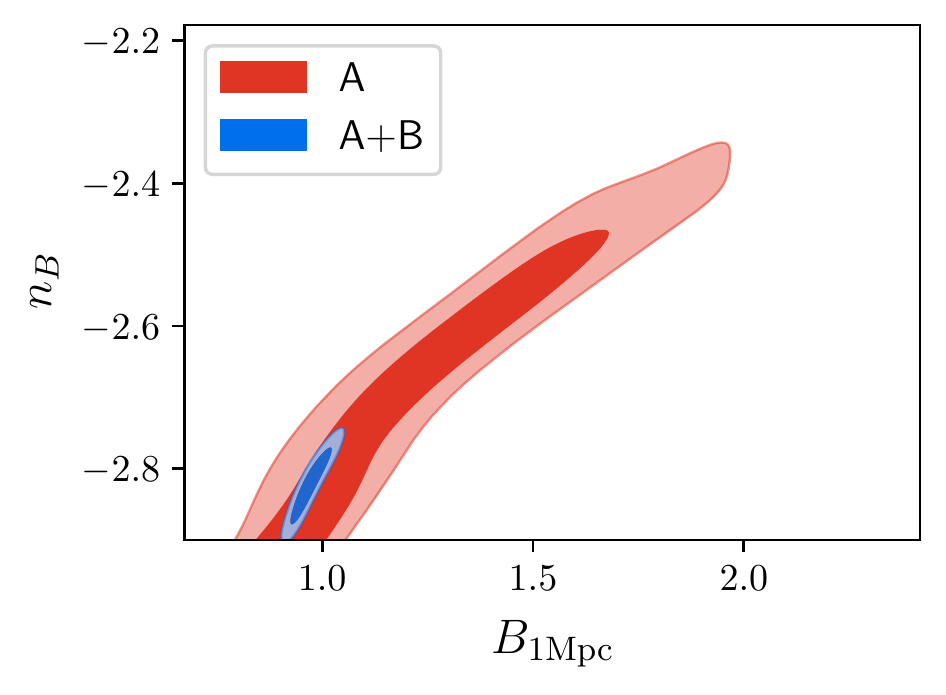}
    \caption{Joint posterior distributions of the magnetic field
      parameters after fitting a \lcdm+PMF model to the simulated CMB
      power spectra with a fiducial model of inflationary tensor modes 
      with $r=0.01$. The red contour
      shows the posterior distribution obtained from Expt~A only,
      while the blue contour shows the posterior distribution as a
      result of a joint constraint from Expt~A and Expt~B, as
      specified in \Tab{tab:3-1}. The levels indicate the 68\% and
      95\% confidence levels, respectively.}
    \label{fig:joint}
\end{figure}

In \Fig{fig:joint}, we show how the joint posterior distribution of
the magnetic field parameters ($B_{1{\rm Mpc}}$ and $n_B$) changes
after we include the data from Expt~B to the constraint. 
The degeneracy between $n_B$ and $\Bmpc$ is broken
when the additional observations from Expt~B (or an SO LAT-like
experiment) are included which tightly constrains the small scale
modes of the primordial magnetic field. The joint constraint leads to
a much tighter allowed parameter space, shown as the red contour, 
favoring a primordial magnetic field
with $\Bmpc \sim 1$\,nG and a scale-invariant spectrum. We find a
$\Delta \chi^2 \simeq -2.5$ between the best-fit \lcdm+r model and the \lcdm+PMF
model, showing a stronger preference to the \lcdm+r model. This
improvement in $\Delta \chi^2$ is driven by the stronger
constraining power in the small angular scales on the compensated
vector-mode signal from primordial magnetic field which dominates at
small angular scales ($\ell \gtrsim 1000$) and has no degenerate signal in
\lcdm+r. This indicates that if an apparent primordial B-mode signal is
detected at an amplitude of around $r=0.01$, a joint constraint using both large and
small angular scale measurements is a promising approach
to rule out a degenerate \lcdm+PMF model.

% ++++++++++++++++++++++++++++++++++++++++++++++++++
% ++++++++++++++++++++++++++++++++++++++++++++++++++

\subsection{Lower $r$ targets} \label{sec:3.2}

In addition to the fiducial model with $r=0.01$ discussed in the
preceding section, we also repeat the study in \Sec{sec:3.1} for
different targets of $r$ ranging from $0.001$ to $0.010$, and compute
$\Delta \chi^2$ between the two best-fit models for each set of the
simulations of a given $r$. In particular, we consider three sets of
combined observations specified as C1, C2, C3 in \Tab{tab:3-1}. C1
represents the set of observations considered in \Sec{sec:3.1} as a
joint constraint of Experiment A and B, C2 represents a similar set of
experiments with lower noise levels, and C3 represents the same set of
experiments in a noise-less limit.

\begin{figure}[t]
  \centering
  \includegraphics[width=0.5\textwidth]{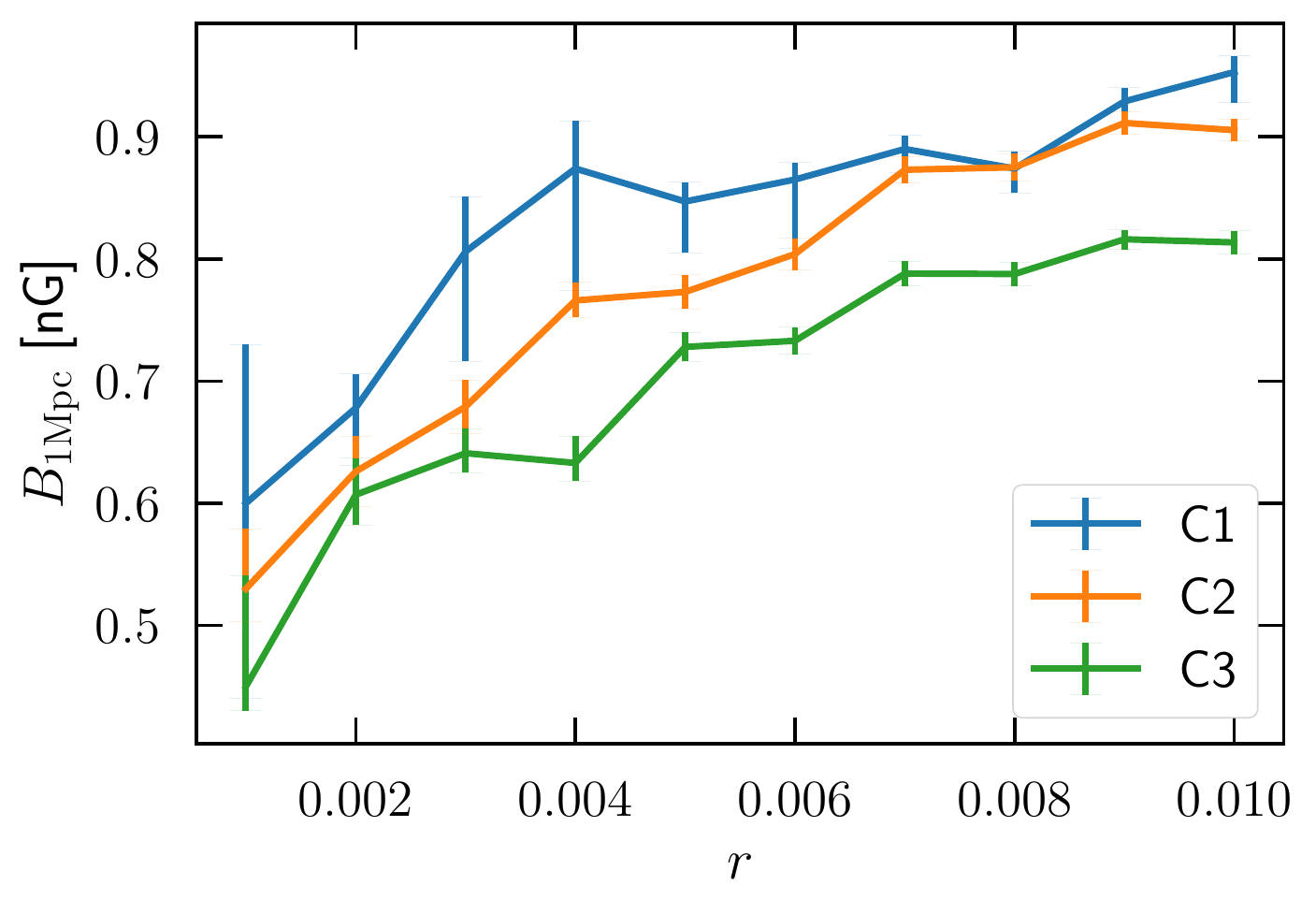}
  \caption{\label{fig:bvr}The magnitudes of magnetic field ($\Bmpc$)
    that fits the simulated data at different target $r$ for
    Expt C1, C2, C3 specified in \Tab{tab:3-1}. The error
    bars indicate the 68\% confidence interval for the marginal
    posterior distribution.}
\end{figure}

The results of model-fitting show that the degenerate \lcdm+PMF models
generally favor a nearly scale-invariant spectrum ($n_B\simeq -2.9$) with
$\Bmpc\lesssim 0.8$\,nG, which is below the current observational limits.
\Fig{fig:bvr} shows how the amplitude of the magnetic field in the
degenerate \lcdm+PMF model varies with $r$. This is useful as it gives
us a reference to what range of the primordial magnetic field
parameter space is of interests to a particular $r$ target. It shows
that, in general, one needs only worry about scale-invariant
primordial magnetic field models with $\Bmpc\gtrsim 0.5$\,nG when targetting
$r\gtrsim 0.001$. The results also show that, as the noise level of the
experiment improves, more magnetic field parameter space will be
strongly constrained, thus reducing the allowed amplitude of the
degenerate primordial magnetic field model.

% ++++++++++++++++++++++++++++++++++++++++++++++++++
% ++++++++++++++++++++++++++++++++++++++++++++++++++

\begin{figure}[t]
  \centering
  \includegraphics[width=0.5\textwidth]{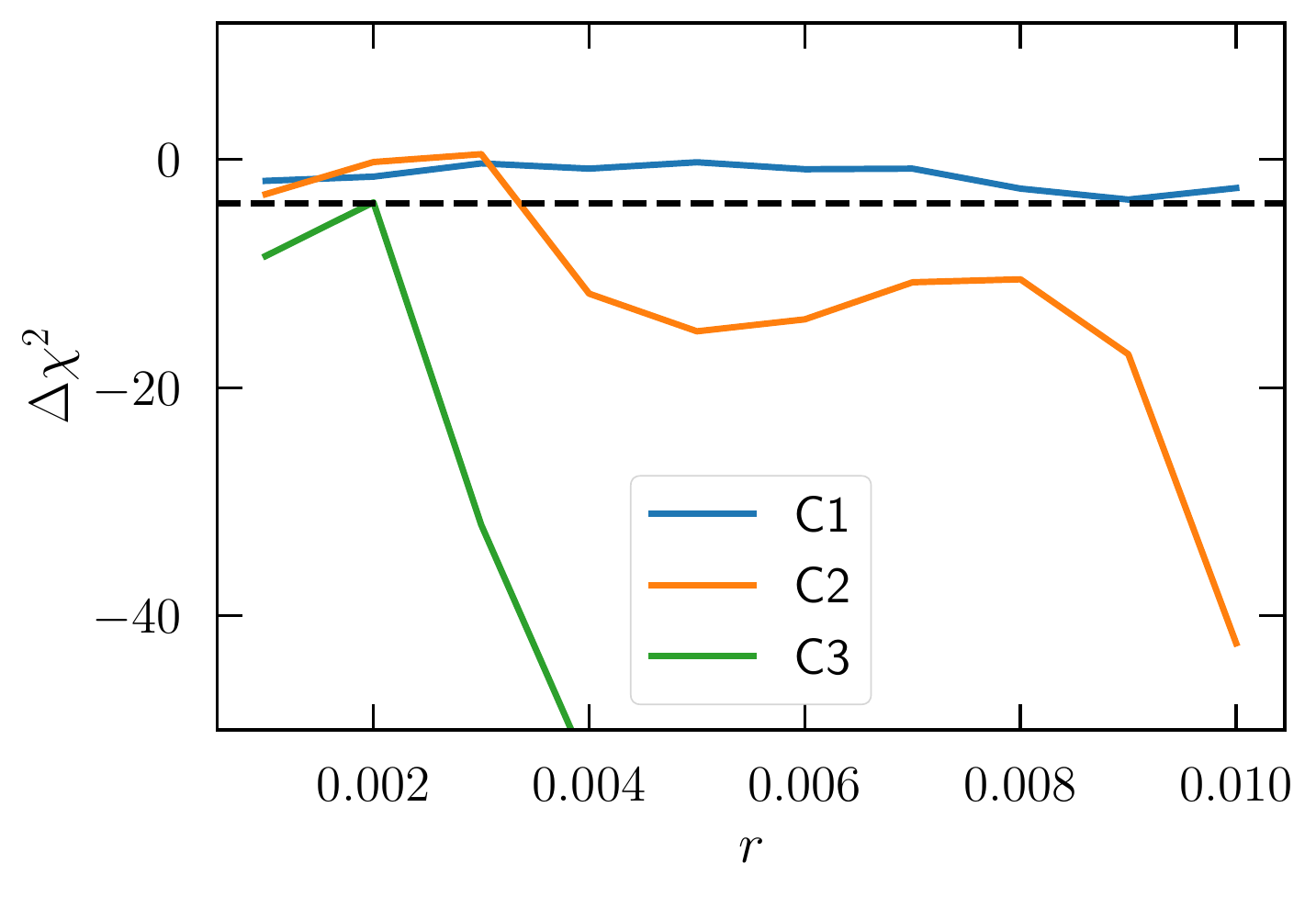}
  \caption{\label{fig:dchi2dr}How $\Delta \chi^2$ varies with different
    targets of $r$. The three lines represent the three simulated set
    of observations specified in \Tab{tab:3-1}. The black dashed
    line shows a reference level of $\Delta\chi^2=-3.841$ which corresponds to
    the 95\% confidence level for a $\chi^2$ distribution with one degree
    of freedom.}
\end{figure}

In \Fig{fig:dchi2dr} we show how $\Delta \chi^2$ between the two best-fit
models changes as we vary $r$ for each of the three sets of simulated
observations. As a reference, we compare the $\Delta \chi^2$ with a 95\%
confidence level of a $\chi^2$ distribution with one degree of freedom
($\Delta \chi^2=-3.841$) since the two competing models differ by one degree
of freedom. We note that the results feature an apparent trend
particularly for Expt~C2, and also some fluctuations particularly at
$r\lesssim 0.004$. This is likely due to a combination of realization-induced
randomness and a poor convergence of some of the MCMC chains.
Nevertheless, combined with \Fig{fig:bvr}, one sees a generic trend in
the reduction of $\Bmpc$ and the increasing of $\Delta\chi^2$ as noise level
reduces or as $r$ is lowered, which matches our expectations. Thus our
results are likely sensible approximations of the future performances,
which are sufficient for our discussion here. In particular, one can
see that the performance of Expt~C1 in breaking the degeneracy between
the two models quickly degrades as $r\lesssim0.008$. With Expt~C2 which has a
much lower noise level similar to the targeting performance of the
CMB-S4 experiments \cite{S4:2016:SciBook}, the situation is much
improved as the degeneracy is effectively broken for any
$r\gtrsim 0.004$. In the noise-less limit (C3), the degeneracy limit is
pushed further down to $r\lesssim 0.002$. This implies that we will be cosmic
variance limited to make a distinction between an inflationary
tensor-mode signal and a primordial magnetic field signal below
$r\lesssim 0.002$.

Note that our conclusions so far are based entirely on constraining
primordial magnetic field through its effects on the CMB power spectra
by means of metric perturbations and Lorentz force. However, this is
not the only way one can constrain primordial magnetic field signals.
In fact, primordial magnetic field also induces a Faraday rotation
effect on the polarization of the CMB photons \citep{Kosowsky_2005},
thus providing an additional means to constrain primordial magnetic
field models. Hence, in the subsequent sections we will examine
whether such effect can improve our ability to distinguish the two
models.

\section{B-Mode Polarization from Faraday Rotation} \label{sec:ch2:faraday}

Another probe of primordial magnetic field is through the effect of Faraday rotation, in which the presence of magnetic field causes a net rotation of the linear 
polarization directions of the CMB photons along their path. The rotation angle $\alpha$ depends on the frequency of observation and the integrated electron density along the line of sight, 
\begin{equation}
  \alpha = \frac{3}{16\pi^2e}\lambda_0^2 \int\dot{\tau}\tilde{\mathbf{B}}(\mathbf{x})\cdot d{\bm l},
\end{equation}
where $\lambda_0$ is the observed wavelength,
$\dot{\tau} \equiv n_e\sigma_T a$ is the differential optical depth proportional to the electron number density $n_e$, 
and $\tilde{\mathbf{B}}\equiv \mathbf{B} a^2$ is the comoving magnetic field.
For a homogeneous magnetic field with a present amplitude of $\sim 1$\,nG, the net rotation of the
polarization angle is about a degree at 30\,GHz, with the
size of the effect scaling with frequency as $\alpha \propto v^{-2}$ \citep{Kosowsky:1996:FR}. For a stochastic magnetic field with a power spectrum $P_B(k)$, the rotation field $\alpha(\nhat)$ is anisotropic with a 2-point correlation function given by \cite{Pogosian_2011}
\begin{equation}
  \begin{split}
  \langle\alpha(\hat{n})&\alpha(\hat{n}')\rangle = \left(\frac{3\lambda_0^2}{16\pi^2 e}\right)^2 \int \frac{d^3 k}{(2\pi)^3}P_B(k)\int d\eta \int d\eta' \\
  &\times \dot{\tau}(\eta)\dot{\tau}(\eta') e^{-i\mathbf{k}\cdot\hat{n} \eta}e^{i\mathbf{k}\cdot \hat{n}'\eta'}[\hat{n}\cdot \hat{n}'-(\hat{k}\cdot \hat{n})(\hat{k}\cdot\hat{n}')],
  \end{split}
\end{equation}
which can also be written as 
\begin{equation}
\langle\alpha(\hat{n})\alpha(\hat{n}')\rangle = \sum_L \frac{2L+1}{4\pi}C_L^{\alpha\alpha}P_L(\hat{n}\cdot\hat{n}'),
\end{equation}
with $P_L(x)$ the Legendre polynomials and $C_L^{\alpha\alpha}$ the
rotational power spectrum. The rotational power spectrum follows as
\begin{equation}
  \label{eq:claa}
C_L^{\alpha\alpha} = \left(\frac{3 \lambda_{0}^{2}}{16 \pi^{2} e}\right)^{2}\frac{2L(L+1)}{\pi}\int_0^\infty \frac{dk}{k} k^3 P_B(k)T_L^2(k),
\end{equation}
where we have defined a transfer function $T_L(k)$ as
\begin{equation}
  \label{eq:Tl}
T_L(k) \equiv \int_0^{k\Delta\eta} \frac{dx}{x} ~\dot{\tau}(\eta_0-x/k)j_L(x).
\end{equation}
Here $\eta_0$ is the conformal time today, $j_L(x)$ is the Spherical
Bessel function, and $\Delta \eta\equiv \eta_0-\eta_*$ with $\eta_*$ corresponding to the conformal time at the maximum visibility. \Eq{eq:claa} provides the general
expression for the rotational power spectrum generated by a primordial magnetic field model with a
given $P_B(k)$.

The rotation field effectively
turns E-mode polarization into B-mode polarization, leading to a B-mode power spectrum $C_\ell^{\rm BB}$ given by \cite{Pogosian_2011}
\begin{equation}
  \label{eq:clbb}
  \begin{split}
    C_{\ell}^{\rm BB} = &\sum_{\ell_2 L}\frac{(2L+1)(2\ell_2+1)}{2\pi} C_L^{\rm \alpha\alpha}C_{\ell_2}^{\rm EE}\left(H_{\ell\ell_2}^L\right)^2\\&~~~\times \left(1+(-1)^{L+\ell+\ell_2}\right),
  \end{split}
\end{equation}
where $H_{\ell\ell_2}^L$ is defined through the Wigner 3j symbol
\citep{book:wigner} as
\begin{equation}
  \label{eq:4.2}
  H_{\ell \ell_2}^L \equiv
  \begin{pmatrix}
    \ell & L & \ell_2\\
    2 & 0 & -2
  \end{pmatrix}.
\end{equation}
\Eq{eq:clbb} gives the expected signal in $\ClBB$ from an anisotropic rotation field $\alpha(\nhat)$ with a power spectrum $C_L^{\alpha\alpha}$, giving us an additional
means to probe the primordial magnetic field model through the Faraday rotation effect.

\subsection{Faraday rotation from a scale-invariant primordial magnetic field}
As discussed in \Sec{sec:3.1}, primordial magnetic field models that generate potentially
degenerate B-mode signals to the primordial gravitational wave are
approximately scale-invariant. Hence we focus exclusively on 
the this class of primordial magnetic field models (with $n_B\simeq-2.9$) in this section. 
In addition, we make
another simplifying assumption that the magnetic modes with scales
smaller than the thickness of the last scattering surface contribute
negligibly to the total Faraday rotation, so we only
consider magnetic modes for $k\lesssim k_D$ with $k_D \simeq 2$\,Mpc$^{-1}$. 
This assumption
is motivated by the fact that the total Faraday rotation is dominated
by the large-scale modes, as the rotation
generated by magnetic modes with scales smaller than the thickness of
the last scattering surface tends to cancel due to the
Faraday depolarization effect \cite{Sokoloff:1998:faraday_depolar}.

\begin{figure}[t]
  \centering
  \includegraphics[width=0.5\textwidth]{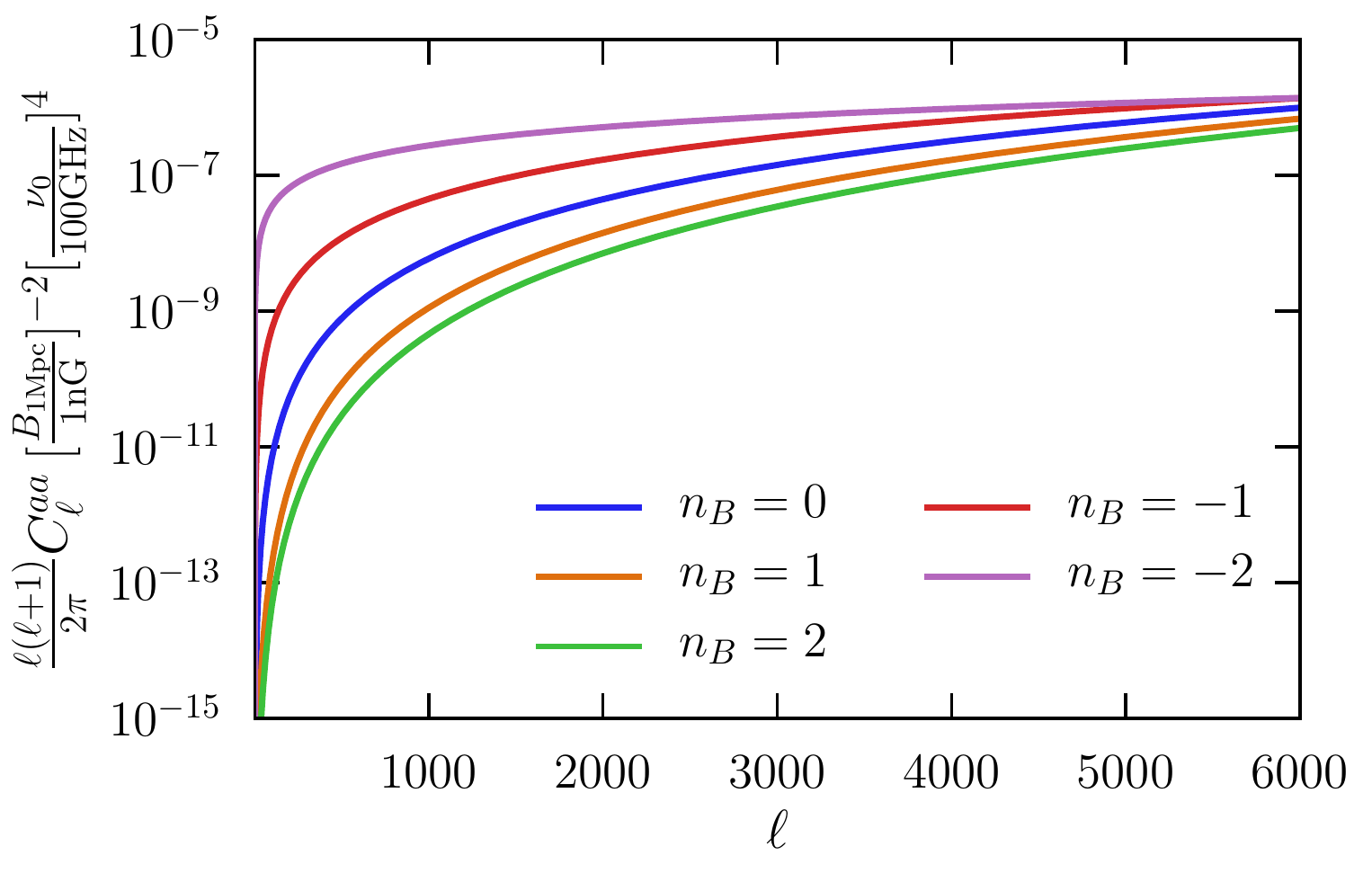}
  \caption{\label{fig:claa}Rotation power spectrum for different
    magnetic spectral indices $n_B$ calculated using \Eq{eq:claa2}
    with the Planck 2018 best-fit cosmology
    \cite{Planck2018:VI:CP}, $\nu_0=100$\,GHz, and $\Bmpc=1$\,nG. The
    amplitude of the power spectrum scales with $\Bmpc^2$ and
    $\nu_0^{-4}$.}
\end{figure}

With the assumptions above, the transfer function $T_L(k)$ defined in
\Eq{eq:Tl} can then be approximated as
\begin{equation}
  T_L(k) \simeq \frac{j_L(k\eta_0)}{k\eta_0},
\end{equation}
where we have used the approximation that $\Delta\eta \approx \eta_0$ and the
fact that the differential optical depth $\dot{\tau}$ is sharply peaked
relative to the slowly varying magnetic field (as we have ignored the
fast varying modes with scales smaller than the thickness of the last
scattering surface) and integrates to $\simeq 1$ near the last 
scattering surface. The rotation power spectrum $C_L^{\rm \alpha\alpha}$ 
then becomes
\begin{equation}
  \label{eq:claa2}
  \begin{split}
    C_L^{\alpha\alpha} &= \frac{9L(L+1)\lambda_0^4}{4(2\pi)^5e^2\eta_0^3}\int_0^{k_D}\mathrm{d}k~ P_B(k)j_L^2(x)\\
    &= \frac{9L(L+1)B_\lambda^2\nu_0^{-4}}{(4\pi)^3\Gamma\left(\frac{n_B+3}{2}\right)e^2}\left(\frac{\lambda}{\eta_0}\right)^{n_B+3}
    \int_0^{x_D}dx~x^{n_B}j_L^2(x),
  \end{split}
\end{equation}
where $x_D \equiv k_D \eta_0$, $\nu_0$ is the observing frequency, and $\lambda=1$\,Mpc is the
length of the smoothing kernel. This
result is consistent with that given in Ref.~\citep{Kosowsky_2005}.
Specifically, we follow the same approximation as in Ref.~\citep{Kosowsky_2005}
that replaces $j_L^2(x)$ with $1/2x^2$ after the second zero of $j_L(x)$ in
\Eq{eq:claa2} to simplify the numerical integration of the fast
oscillating functions. In \Fig{fig:claa} we show the rotation power spectrum
of a primordial magnetic field with $\Bmpc=1$\,nG for different $n_B$, as calculated from \Eq{eq:claa2}. 
The results show that as the spectral index
approaches $n_B \simeq -3$, the rotation spectrum approaches a
scale-invariant limit as expected. The above derivations 
assume the CMB polarization is generated instantaneously in the beginning of
recombination, which is not true. A full calculation also needs to consider 
that Faraday rotation occurs alongside with the generation of CMB 
polarization. This effect has been calculated in 
Ref.~\citep{Pogosian_2011} and shown to result in difference small compared to
our order of magnitude estimate here. 

% ++++++++++++++++++++++++++++++++++++++++++++++++++++++++++++++++++++
% claa plot for different nB
% ++++++++++++++++++++++++++++++++++++++++++++++++++++++++++++++++++++

\begin{figure}[t]
  \centering
  \includegraphics[width=0.45\textwidth]{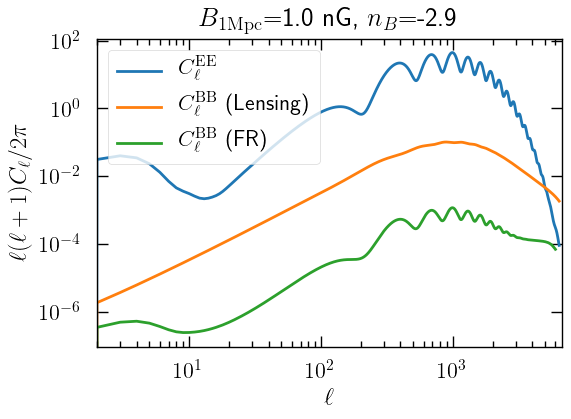}
  \caption{\label{fig:clbb}The green curve shows the B-mode signal (in units of $\mu$K$^2$)
    generated by the Faraday rotation of a primordial magnetic field with $n_B=-2.9$ and
    $\Bmpc=1$\,nG at $\nu_0=100$\,GHz. The orange curve shows the
    expected lensing signal and the blue curve shows the
    $C_\ell^{\rm EE}$ signal. Note that ``FR'' denotes Faraday rotation.}
\end{figure}

With the rotational power spectrum $C_\ell^{\alpha\alpha}$, one can then estimate
the expected $\ClBB$ power spectrum sourced by the rotation field
using \Eq{eq:clbb}. In \Fig{fig:clbb}, we show the expected B-mode
power spectrum sourced by a nearly scale-invariant primordial magnetic field with $n_B=-2.9$
and $\Bmpc=1$\,nG, observed at 100\,GHz. The result shows two noticeable features: (1)
Faraday rotation signal in $\ClBB$ peaks at small angular scales
(at $\ell \sim 1000$), similar to the CMB lensing signal, with a
significantly lower amplitude than CMB lensing; (2) Unlike the CMB lensing
signal, the B-mode signal from the rotation field displays acoustic
oscillations similar to those in CMB E-mode power spectrum. This is
expected since, according to \Eq{eq:clbb}, the B-mode signal from the
rotation field is effectively a convolution of the E-mode power
spectrum $\ClEE$ with the rotation power spectrum $\Claa$ in
$\ell$-space. $\Claa$ is scale invariant, so the variation with $\ell$ in the
resulting $\ClBB$ is determined by that of $\ClEE$, thus reflecting the
acoustic oscillations. This is a unique feature that allows 
distinguishing the rotation signal from the lensing signal in the
$\ClBB$.

To project the performance of future CMB experiments in constraining the
primordial magnetic field by measuring the Faraday rotation signal, we define the
signal-to-noise ratio (SNR) as
\begin{equation} \label{eq:clbb-snr}
  \left(\frac{S}{N}\right)^2 = \sum_\ell\frac{(2\ell+1)f_{\rm sky}\left(C_\ell^{\rm BB,FR}\right)^2}{2\left(C_\ell^{\rm BB,tot} + N_\ell^{\rm BB}\right)^2},
\end{equation}
with $C_\ell^{\rm BB,FR}$ the expected B-mode signal from the Faraday
rotation, and $C_\ell^{\rm BB,tot}$ the total B-mode signal that includes
the contributions both the Faraday rotation signal and the CMB lensing
signal. $N_\ell^{\rm BB}$ refers to the expected B-mode noise power
spectrum from a given experiment as approximated by \Eq{eq:ch2:nl}.
The factor $\fsky$ is added to approximate the effect of the partial
sky coverage of a realistic experiment, in the form of a reduction in
the number of available measurements and thus a reduction in the total
SNR. In addition, we assume an observing frequency of 100\,GHz for the
subsequent discussion. Lower frequencies increase the rotation signal for
a given magnetic field, but are also technically more difficult to attain
comparable map sensitivity and resolution.

% ++++++++++++++++++++++++++++++++++++++++++++++++++++++++++++++++++++
% clbb
% ++++++++++++++++++++++++++++++++++++++++++++++++++++++++++++++++++++

\begin{figure}[t]
  \centering
  \includegraphics[width=0.45\textwidth]{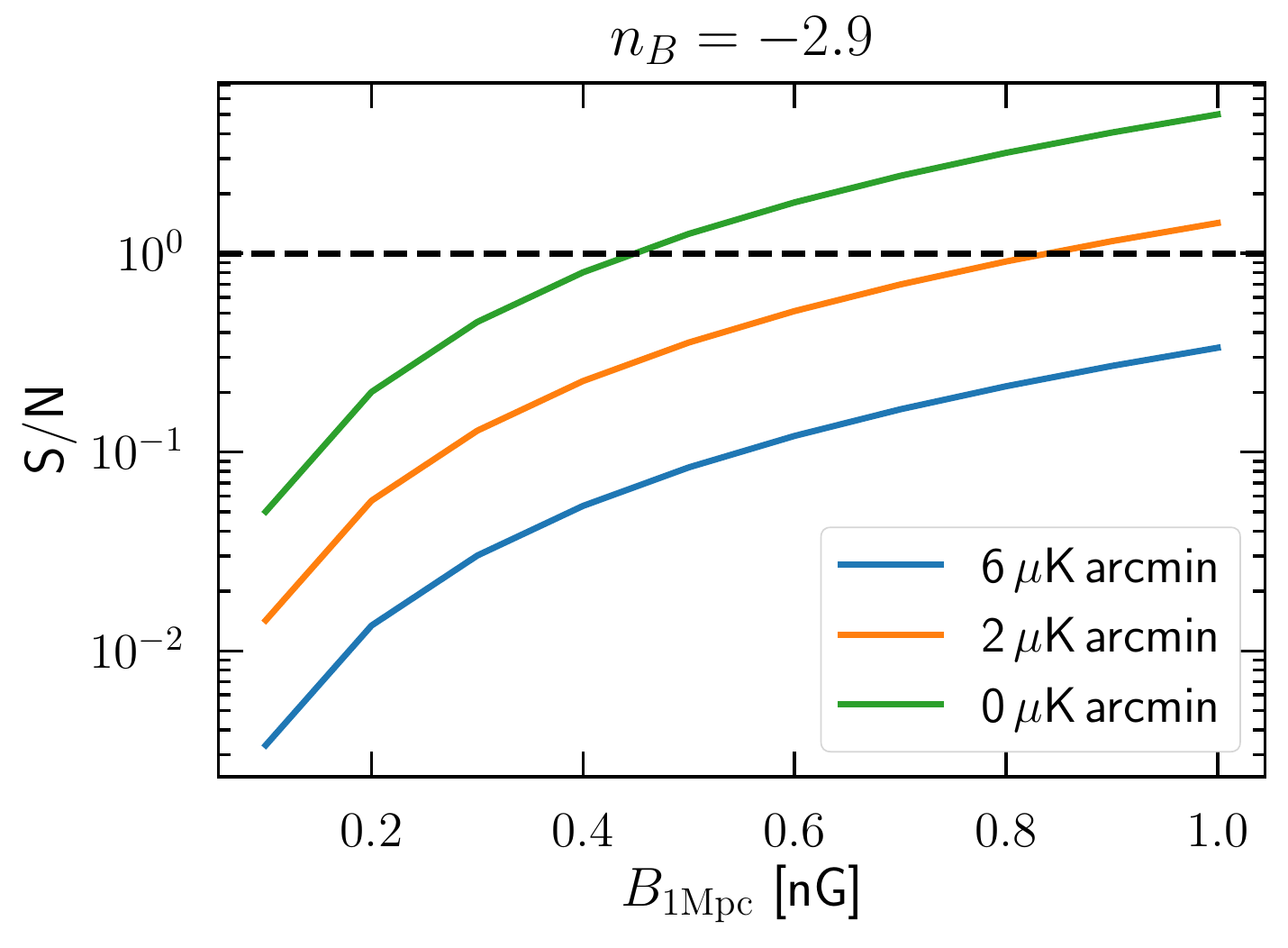}
  \caption{\label{fig:snr-clbb} Signal-to-noise ratio for various
    $\Bmpc$. The three different solid curves show the S/N curve for
    three experiments with various noise levels. The dashed curve
    indicates the threshold of $S/N=1$.}
\end{figure}

As the Faraday rotation signal is significant mainly on small
angular scales, large-aperture experiments are most relevant to detecting such signal. 
Specifically, we consider Expt B as specified in
\Tab{tab:3-1} with different noise levels (6\,$\mu$K\,arcmin,
2\,$\mu$K\,arcmin, and 0\,$\mu$K\,arcmin), and compute the SNR for each
experiment for a scale-invariant primordial magnetic field with the amplitude $\Bmpc$
varying from 0.1\,nG to 1\,nG. The resulting SNRs are presented in
\Fig{fig:snr-clbb}, which shows that for an SO LAT-like experiment
with a noise level of 6\,$\mu$K\,arcmin, the Faraday rotation signal is
not detectable in the power spectrum, hence contributing negligible
constraining power on the primordial magnetic field. In comparison, a CMB S4-like experiment
with a noise level of 2\,$\mu$K\,arcmin barely detects a primordial magnetic field with
$\Bmpc\gtrsim 0.9$\,nG at SNR\,$\gtrsim 1$, while at the noiseless limit, one can
detect a primordial magnetic field with $\Bmpc\gtrsim 0.5$\,nG with SNR\,$\gtrsim 1$, and
$\Bmpc\gtrsim 0.8$\,nG with SNR\,$\gtrsim 3$. As concluded from \Fig{fig:bvr}, degenerate
primordial magnetic field models of interest to the upcoming experiments generally have
amplitudes $\Bmpc$ ranging from $\sim 0.5-1$\,nG, comparable to the
detection limit of the noiseless case. This suggests that Faraday
rotation in the B-mode power spectrum is unlikely a competitive
constraint on the primordial magnetic field.
% Our results also show that as we approach the noise-less
% limit, we can constrain primordial magnetic field down to $B\gtrsim0.3$\,nG with S/N$>$1.

% ++++++++++++++++++++++++++++++++++++++++++++++++++++++++++++++++++++
% claa plot for different nB
% ++++++++++++++++++++++++++++++++++++++++++++++++++++++++++++++++++++

On the other hand, the above SNR estimates
neglect the effect of delensing, which is a procedure to
remove the CMB lensing signal from the B-mode power spectrum (see,
e.g., \citep{Smith:2012:delens}). As the CMB lensing signal is
generally much larger than the Faraday rotation signal
in $\ClBB$, being able to remove a significant portion of the lensing
signal significantly reduces the total variance in the B-mode power
spectrum, thus improving the SNR. To be more specific, we can denote
the $C_\ell^{\rm BB,tot}$ in \Eq{eq:clbb-snr} as
\begin{equation}
C_l^{\rm BB,tot} = C_l^{\rm BB,CMB} + C_l^{\rm BB,FR} + A_{\rm delens}C_\ell^{\rm BB,lensing},  
\end{equation}
where $ C_l^{\rm BB,CMB}$, $C_\ell^{\rm BB,FR}$, and
$C_\ell^{\rm BB,lensing}$ denote the B-mode signal from the CMB,
primordial magnetic field, and lensing, respectively, and
$A_{\rm delens}$ denotes the residual fraction of delensing which
characterizes the delensing efficiency. Optimistic estimates suggest
that an SO-like experiment can achieve $A_{\rm delens}\sim 0.5$ with
inputs from external datasets \citep{SO:2019:SciGoal}, and a CMB
S4-like experiment with a noise level around 2\,$\mu$K\,arcmin can
achieve $A_{\rm delens}\sim 0.4$ \cite{S4:2016:SciBook}. If the B-mode power
spectrum is signal dominated, delensing can improve the
signal-to-noise ratio by a factor of $A_{\rm delens}^{-1}$, thus
lowering the primordial magnetic field detection limit by a factor of
$A_{\rm
  delens}^{-1/2}$.

\section{Rotational field reconstruction from primordial magnetic field} \label{sec:ch2:rot}
Faraday rotation acts as an effective rotation field
$\alpha(\nhat)$ that rotates the CMB polarization field:
\begin{equation}
{}_{\pm 2}A(\hat{\mathbf{n}}) \equiv (Q\pm iU)(\hat{\mathbf{n}}) = e^{\pm 2i\alpha(\hat{\mathbf{n}})}(\tilde{Q}\pm i\tilde{U})(\hat{\mathbf{n}}),
\end{equation}
where $Q$ and $U$ refer to the Stoke parameters for the rotated polarization field
and we use tilde to denote the unrotated polarization field. In the limit that
$\alpha(\hat{\mathbf{n}}) \ll 1$,
$\delta_{\pm 2}A(\nhat) \simeq \pm 2i\alpha(\nhat){}_{\pm
  2}\tilde{A}(\nhat)$. 
Such rotation induces off-diagonal correlations between E-mode 
and B-mode polarization maps \citep{Kamionkowski_2009,2009PhRvD..79l3009Y} (see Appendix~\ref{app:qe} for a derivation), 
given by
\begin{equation}
  \label{eq:eb}
  \langle E_{lm}B_{l'm'}^*\rangle_{\rm CMB} = ~\sum_{LM}\alpha_{LM}\xi_{lml'm'}^{LM}f_{lLl'}^{\rm EB},
\end{equation}
with
\begin{equation}
  f_{lLl'}^{\rm EB} = 2\epsilon_{lLl'}[H_{l'l}^L\tilde{C}_l^{\rm EE}-H_{ll'}^L\tilde{C}_l^{\rm BB}],
\end{equation}
\begin{equation}
\begin{split}
\xi_{lml_2m_2}^{LM} \equiv (-1)^m &\sqrt{\frac{(2L+1)(2l_2+1)(2l+1)}{4\pi}}\\
&\qquad\times\begin{pmatrix}
  l & L & l_2\\
  -m & M & m_2
\end{pmatrix},
\end{split}
\end{equation}
and
\begin{equation}
  \epsilon_{l L l_2} \equiv \frac{1+(-1)^{l+L+l_2}}{2}.
\end{equation}

% ++++++++++++++++++++++++++++++++++++++++++++++++++++++++++++++++++++
\begin{figure}[t]
  \centering
  \includegraphics[width=0.5\textwidth]{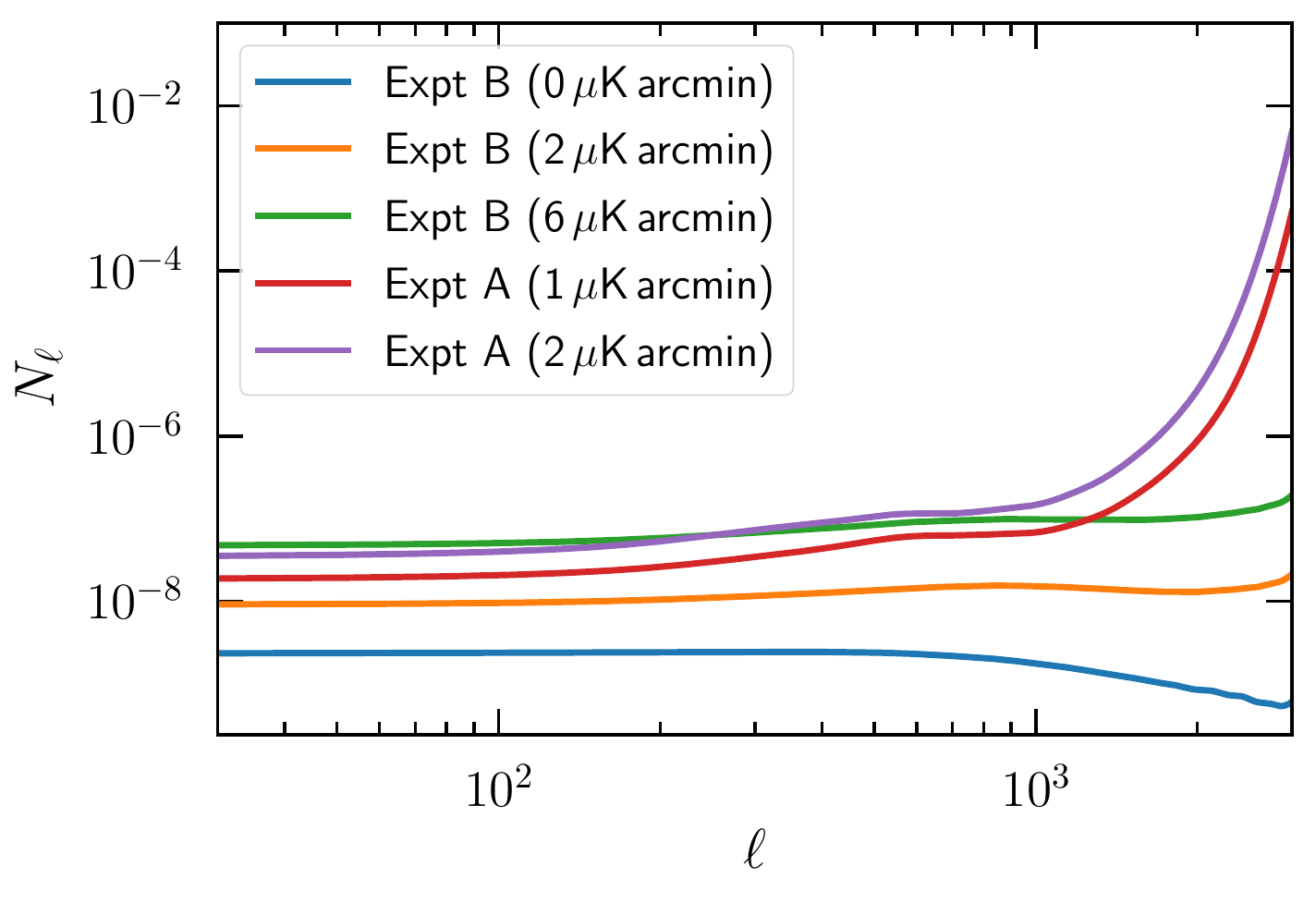}
  \caption{Gaussian noise covariance $N_L^{\rm EB}$ for experiments
    specified in \Tab{tab:3-1} with varying noise levels.}
  \label{fig:5.1}
\end{figure}
% ++++++++++++++++++++++++++++++++++++++++++++++++++++++++++++++++++++

The $\langle...\rangle_{\rm CMB}$ denotes that the average is to be
taken over CMB realisations only. The coupling also allows one to
reconstruct the rotation field $\alpha_{LM}$ with a quadratic estimator
similar to the reconstruction of CMB lensing \cite{Hu:2001:lensflat}:
\begin{equation}
\hat{\alpha}_{LM} = A_L^{\rm EB}\sum_{ll'}\sum_{mm'}\xi_{lml'm'}^{LM}g_{ll'}^{\rm EB}E_{lm}B_{l'm'}^*,
\end{equation}
with normalization factor $A_L$ defined as
\begin{equation}
  \left(A_L^{\rm EB}\right)^{-1} = \sum_{ll'}\frac{(2l+1)(2l'+1)}{4\pi}g_{ll'}^{\rm EB}f_{lLl'}^{\rm EB},
\end{equation}
ensuring the quadratic estimator is unbiased. The weights
$g_{ll'}^{\rm EB}$ can be chosen to minimize the total variance of the
estimator $\langle\alpha_{LM}^*\alpha_{LM}\rangle$ with
\begin{equation}
g_{ll'}^{\rm EB} = \frac{f_{lLl'}^{\rm EB}}{C_l^{\rm EE}C_{l'}^{\rm BB}},
\end{equation}
The minimized variance of estimator, denoted as $N_L^{\rm EB}$, is
related to the normalization factor as
\begin{equation} \label{eq:ch2:nleb}
  N_L^{\rm EB} = A_L^{\rm EB} = \sum_{ll'}\frac{(2l+1)(2l'+1)}{4\pi}\frac{(f_{lLl'}^{\rm EB})^2}{C_l^{\rm EE}C_{l'}^{\rm BB}},
\end{equation}
with $C_l^{\rm EE}$ and $C_{l'}^{\rm BB}$ the observed E- and B-mode
power spectrum, respectively. Here $N_L^{\rm EB}$ is a dimensionless
quantity that characterizes the variance of the reconstructed rotation
angle at each $L$.

In \Fig{fig:5.1}, we show the expected reconstruction noise $N_L^{EB}$
calculated using \Eq{eq:ch2:nleb} for experiments considered
previously in \Tab{tab:3-1}, and for a nearly scale invariant primordial magnetic field with
varying amplitudes of $\Bmpc$ and $n_B=-2.9$. In particular, we
consider Expt~A with noise levels of 2\,$\mu$K\,arcmin and
1\,$\mu$K\,arcmin, and Expt~B with noise levels of 6\,$\mu$K\,arcmin,
2\,$\mu$K\,arcmin, and 0\,$\mu$K\,arcmin. The results show that the large-aperture experiments have orders of magnitude lower reconstruction
noise at $\ell\gtrsim 1000$, confirming our expectation that the small-scale
CMB anisotropies have stronger constraining power on the Faraday
rotation signal.

% ++++++++++++++++++++++++++++++++++++++++++++++++++++++++++++++++++++
\begin{figure}[t]
  \centering
  \includegraphics[width=0.45\textwidth]{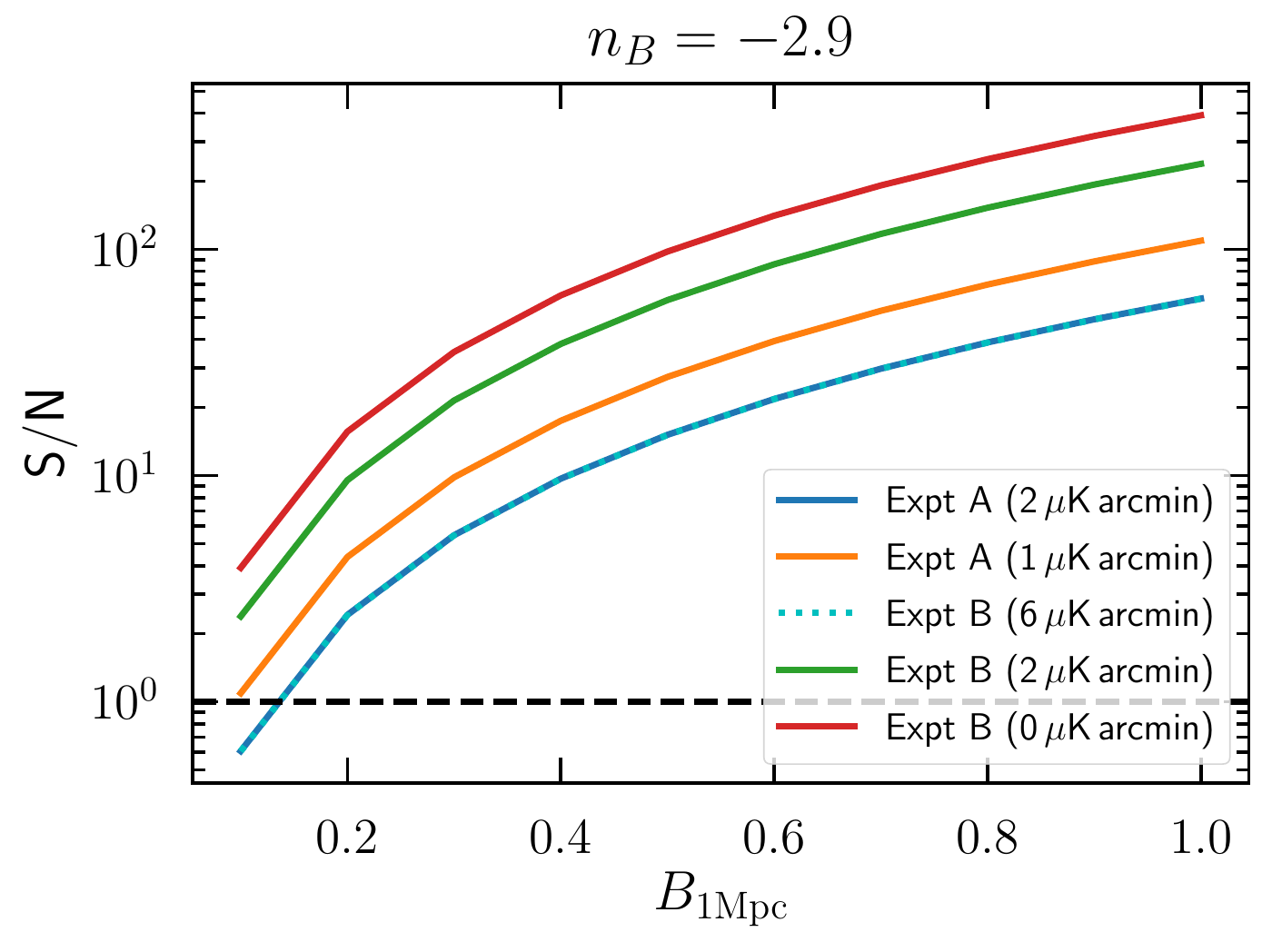}
  \caption{Signal-to-noise ratio expected for the quadratic estimator
    in a variety of experimental settings. The black dashed line
    represents S/N=1.}
  \label{fig:5.2}
\end{figure}
% ++++++++++++++++++++++++++++++++++++++++++++++++++++++++++++++++++++

To forecast the expected performance of the quadratic estimator for
future CMB experiments, we define the SNR as
\begin{equation}
  (S/N)^2 = \sum_{L=L_{\rm min}}^{L_{\rm max}}f_{\rm sky}\frac{2L+1}{2}\left(\frac{C_L^{\alpha\alpha}}{N_L^{EB}}\right)^2,
\end{equation}
where, similar to \Sec{sec:ch2:faraday}, we use $\fsky$ to approximate
the partial sky coverage. We also assume the observations are made at
100\,GHz, which is the frequency channel expected to contribute the highest SNR.

In \Fig{fig:5.2} we show the expected SNR for the same set of
experiments considered previously. It shows that reconstructing a
rotation field using the quadratic estimator approach results in an
order of magnitude improvement in the SNR as compared to constraining
its effects on the CMB B-mode power spectrum. This is consistent
with the claims in \citep{Mandal:2022} and is unsurprising as
the effect of a rotation field $\alpha$ on $C_\ell^{\rm BB}$ scales
as $\alpha^2$, which is a second order effect, whereas its
effect on the cross-correlation $\expect{EB}$ scales with $\alpha$
(see \Eq{eq:eb}), which is a first order effect, thus giving a
significantly improved SNR. The results also show that large-aperture
experiments (Expt B) have better SNR in general as a result of the
significantly lower reconstruction noise (as shown in
\Fig{fig:5.1}). Specifically, a SO SAT-like experiment with a noise
level of 2\,$\mu$K\,arcmin gives comparable SNR to an SO LAT-like
experiment with a noise level of 6\,$\mu$K\,arcmin, both of which are
capable of constraining primordial magnetic field models down to
$\Bmpc\gtrsim0.3$\,nG with $S/N\gtrsim 3$. CMB S4-like noise levels
push down this limit to
$\Bmpc\gtrsim 0.1$\,nG. 

These calculations demonstrate that primordial
magnetic field models which have B-mode power spectra degenerate
with primordial tensor modes with $r>0.001$
will be strongly constrained by the rotation signal in
small-scale anisotropies. As in
Sec.~\ref{sec:ch2:faraday}, we have neglected the effect of delensing,
which may further improve the primordial magnetic field constraint.

\section{Discussion} \label{sec:ch2:discussion}
We have investigated the following question: can primordial magnetic
fields be distinguished from primordial
gravitational waves as a source of B-mode polarization in the CMB power spectrum? Concerns over a
possible degeneracy in the B-mode power spectrum signal have previously been raised (see, e.g., 
\citep{Pogosian:2018:PMF,Renzi_2018}). In this work we have confirmed
with simulations that the answer is likely ``no'' if one utilizes only the
information in the large-scale CMB anisotropies ($\ell\lesssim 1000$), 
as a primordial magnetic field also introduces large-scale B-mode signals 
by sourcing tensor-mode metric perturbations in a mathematically equivalent
form to that of the primordial gravitational waves, thus generating a
completely degenerate signal on large angular scales. However, as we
have further demonstrated, after including 
small-scale CMB polarization anisotropies ($\ell \gtrsim 1000$), the answer becomes
``yes'' because of both the magnetic field's vector-mode contribution to
B-mode polarization on small scales, and especially due to Faraday rotation
of E-mode polarization into B-mode. Upcoming high-sensitivity measurements of
polarization at small scales will enable this distinction for any magnetic field
which might be mistaken for a primordial tensor mode signal when using only
large-angle B-mode polarization data. We have demonstrated this explicitly for
tensor-mode amplitudes down to $r=10^{-3}$. For even smaller tensor-mode signals,
at some point sufficient delensing techniques must be demonstrated. The amplitude
at which lensing signals become an important consideration remains to be seen
\cite{cai:2022:rotbias}.

Our analysis extends previous work (e.g., Refs.~\citep{Renzi_2018}) in considering a wider class of magnetic field models and tensor-to-scalar ratio targets, and more importantly, also in explicitly identifying degenerate magnetic field models to a given tensor-mode signal using simulations and MCMC-based model-fitting.
We also for the first time consider map-based Faraday rotation estimation as a way to break the degeneracy between tensor modes and magnetic fields. 
Our result provides a practical recipe to follow: should a potential tensor-mode signal be detected in the CMB B-mode power spectrum, one can identify degenerate magnetic field models from our analysis and look for its Faraday rotation signal. Upper limits on such a signal provides a clear route to ruling out a plausible contaminant to a tensor B-mode signal. Magnetic fields thus join gravitational lensing and galactic foregrounds as known B-mode contributors for which we possess clear methods of discriminating them from
the hallmark signature of early-universe inflation. 

\section{Acknowledgments}
We thank Daniel Boyanovsky for helpful comments on the manuscript. YG
acknowledges the partial support of PITT PACC in the duration of this
work. This research was supported in part by the University of
Pittsburgh Center for Research Computing through the resources
provided.

\bibliographystyle{unsrtnat}
\bibliography{references}

\clearpage
\onecolumngrid
\appendix
\section{MCMC}
\label{app:post}
We perform MCMC-based model fitting using an ensemble sampler from
{\texttt emcee} \citep{Foreman_Mackey_2013} with 50 walkers. We use
with a mixed proposal function that makes stretch moves 95\% of the
time and Gaussian moves based on the fisher matrix 5\% of the time. We
find that the resulting MCMC chains generally converge well after 400
steps based on autocorrelation tests and adopt a fixed number of 400
steps for all subsequent MCMC runs. Specifically, for \lcdm+r model,
we adopt flat priors on $\omega_b$, $\omega_{\rm CDM}$, $H_0$, $n_s$,
$A_s$, $r$, and a Gaussian prior on $\tau_{\rm reio}$ with
$\tau_{\rm reio} = 0.065\pm0.015$. For the \lcdm+PMF model, we use a flat
prior on $\Bmpc$ with $0\le \Bmpc \le 2.5$\,nG, and a flat prior on
$n_B$ restricted to $-2.9\le n_B\le 0$.

\begin{figure*}[htbp]
\centering
\includegraphics[width=0.9\textwidth]{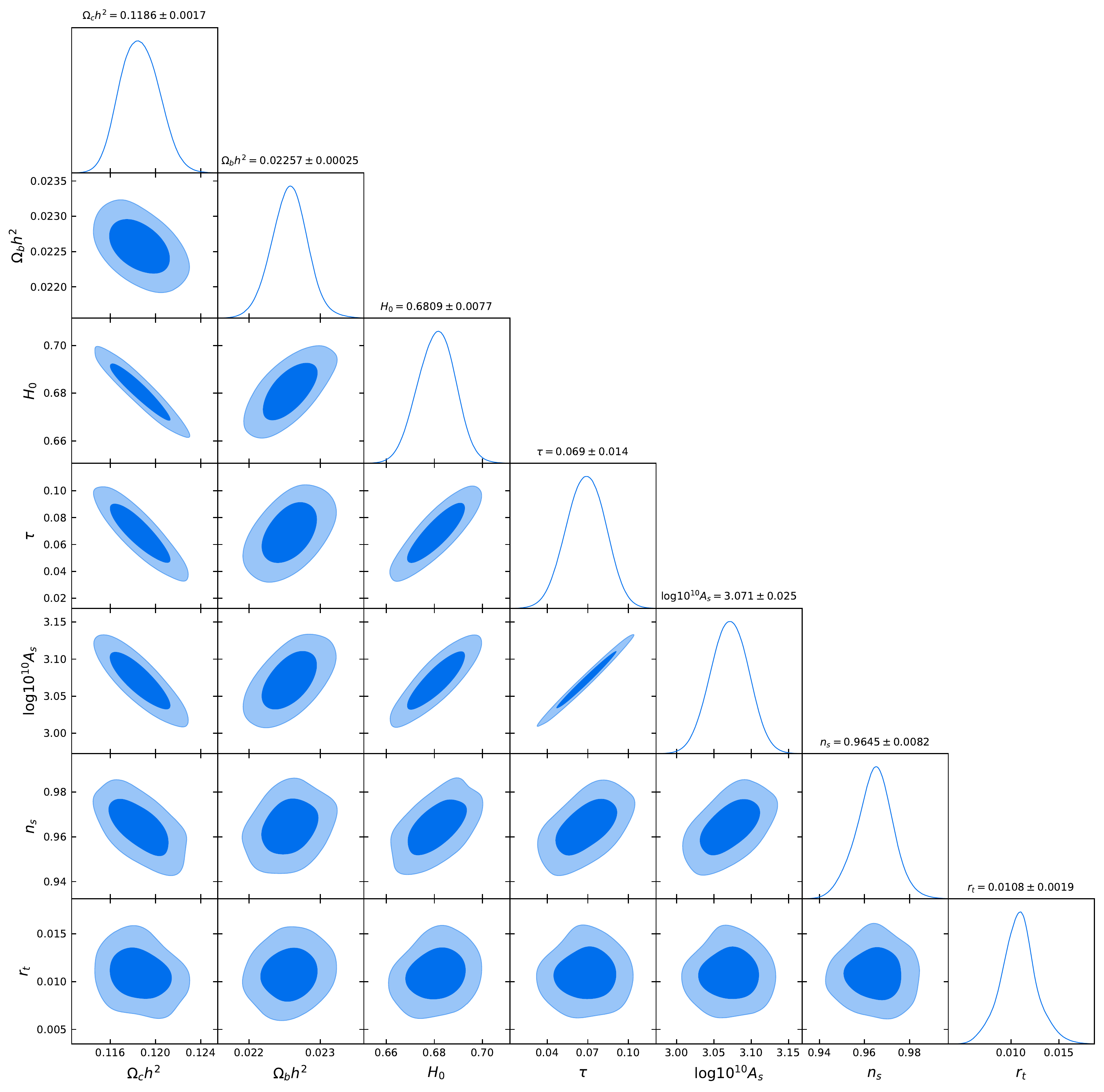}
\caption{Best-fit \lcdm+r cosmological parameters obtained for simulated data for Expt~A with a fiducial cosmology with $r=0.010$.}
\label{fig:post-lcdm+r}
\end{figure*}

\begin{figure*}[htbp]
\centering
\includegraphics[width=0.9\textwidth]{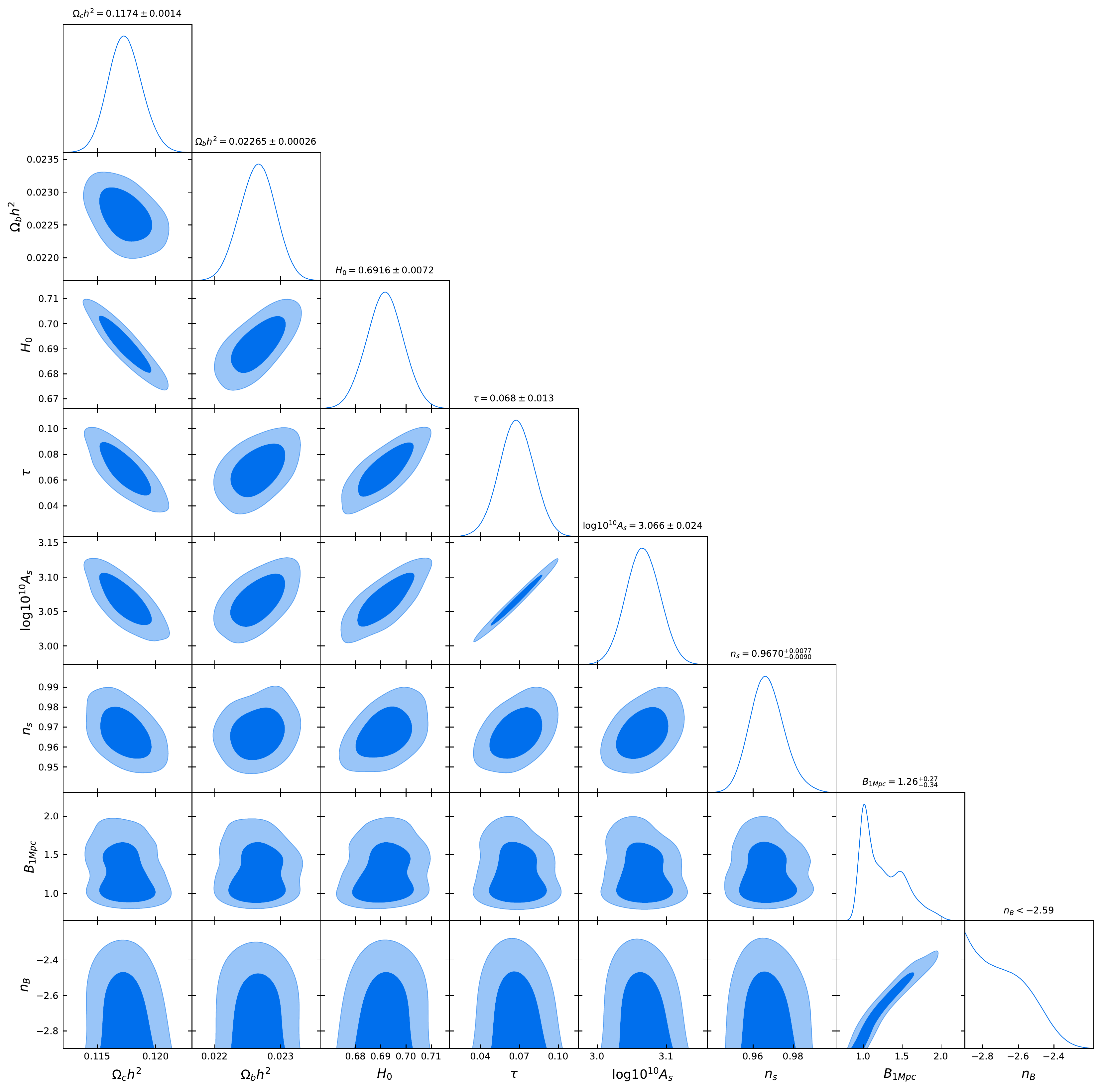}
\caption{Best-fit \lcdm+PMF cosmological parameters obtained for simulated data for Expt~A with a fiducial cosmology with $r=0.010$.}
\label{fig:post-lcdm+pmf}
\end{figure*}

% show full posterior
In Fig.~\ref{fig:post-lcdm+r} and Fig.~\ref{fig:post-lcdm+pmf}, we show the full set of posterior distributions for the \lcdm+r and \lcdm+PMF models respectively, when fitting the simulated observations from Expt~A with a fiducial cosmology with non-zero tensor-to-scalar ratio, $r=0.01$. A burn-in ratio of 70\% has been applied to obtain the posterior distributions.

% ++++++++++++++++++++++++++++++++++++++++++++++++++++++++++++++++++++
% posterior plot of parameter fitting for lcdm+r model
% ++++++++++++++++++++++++++++++++++++++++++++++++++++++++++++++++++++

% ++++++++++++++++++++++++++++++++++++++++++++++++++++++++++++++++++++
% posterior plot of parameter fitting for lcdm+r model
% ++++++++++++++++++++++++++++++++++++++++++++++++++++++++++++++++++++

\section{Quadratic estimator for polarization rotation}\label{app:qe}
Faraday rotation acts as an effective rotation field $\alpha(\nhat)$ which
rotates the CMB polarization maps, given by 
\begin{equation}
{}_{\pm 2}A(\hat{\mathbf{n}}) \equiv (Q\pm iU)(\hat{\mathbf{n}}) = e^{\pm 2i\alpha(\hat{\mathbf{n}})}(\tilde{Q}\pm i\tilde{U})(\hat{\mathbf{n}}),
\end{equation}
where $Q$ and $U$ refer to the Stoke parameters for the rotated CMB
photons. Approximating $\alpha$ as a small angle, the change in the polarization 
field due to rotation can be approximated as
$\delta({}_{\pm 2}A(\nhat)) \simeq \pm 2i\alpha(\nhat){}_{\pm 2}\tilde{A}(\nhat)$. In $lm$ space, the change in $_{\pm 2}A_{lm}$ is
\begin{equation}
  \begin{split}
    \delta({}_{\pm 2}A_{l m}) \simeq& \pm 2i\sum_{\rm LM}\sum_{l_2m_2}\alpha_{LM}{}_{\pm 2}A_{l_2m_2} \\
    &\times \int d\hat{\mathbf{n}}~{}_{\pm 2}Y_{l m}^*(\hat{\mathbf{n}}) Y_{LM}(\hat{\mathbf{n}}){}_{\pm 2}Y_{l_2 m_2}(\hat{\mathbf{n}}),
  \end{split}
\end{equation}
where ${}_sY_{l m}$ denotes the spin-weighted spherical
harmonics \citep{Goldberg:1967:spin}. The integral can be performed with
the formula
\begin{equation}
\begin{split}
\int &d \hat{\mathbf{n}}_{s_{1}} Y_{l_1 m_1}(\hat{\mathbf{n}})_{s_{2}} Y_{l_{2} m_{2}}(\hat{\mathbf{n}})_{s_{3}} Y_{l_{3} m_{3}}(\hat{\mathbf{n}}) = \left[\frac{\prod_{i=1}^32l_i+1}{4\pi}\right]^{1/2}\\
&\times \left(\begin{array}{ccc}
j_{1} & j_{2} & j_{3} \\
m_{1} & m_{2} & m_{3}
\end{array}\right)\left(\begin{array}{ccc}
j_{1} & j_{2} & j_{3} \\
-s_{1} & -s_{2} & -s_{3}
\end{array}\right),
\end{split}
\end{equation}
which gives
\begin{equation}
  \delta({}_{\pm 2}A_{l m}) \simeq \pm 2i\sum_{\rm LM}\sum_{l_2m_2}\alpha_{LM}{}_{\pm 2}A_{l_2m_2}\xi_{lml_2m_2}^{LM}{}_{\pm}H_{ll_2}^L,
\end{equation}
with
\begin{equation}
\xi_{lml_2m_2}^{LM} \equiv (-1)^m\sqrt{\frac{(2L+1)(2l_2+1)(2l+1)}{4\pi}}
\begin{pmatrix}
  l & L & l_2\\
  -m & M & m_2
\end{pmatrix},
\end{equation}
and
\begin{equation}
_\pm H_{ll_2}^L \equiv
\begin{pmatrix}
  l & L & l_2 \\
  \pm 2 & 0 & \mp 2
\end{pmatrix} = (-1)^{l+L+l_2}{}_\mp H_{ll_2}^L.
\end{equation}
On the other hand, the polarization field $_{\pm 2}A_{lm}$ can be decomposed into the
curl-free (E-mode) and the gradient-free (B-mode) components with
\begin{equation}
  \begin{split}
    E_{lm} &= \frac{1}{2}\left(_{+2}A_{lm} + {}_{-2}A_{lm}\right),\\
    B_{lm} &= \frac{1}{2i}\left(_{+2}A_{lm} - {}_{-2}A_{lm}\right).\\
  \end{split}
\end{equation}
This gives
\begin{equation}
  \label{eq:sec5elm}
  \delta E_{lm} = -2i\sum_{LM}\sum_{l_2m_2}\alpha_{LM}\xi_{lml_2m_2}^{LM}H_{ll_2}^L\left(\beta_{lLl_2} \tilde{E}_{l_2m_2} + \epsilon_{lLl_2}\tilde{B}_{l_2m_2}\right),
\end{equation}
and
\begin{equation}
  \label{eq:sec5blm}
  \delta B_{lm} = 2\sum_{LM}\sum_{l_2m_2}\alpha_{LM}\xi_{lml_2m_2}^{LM}H_{ll_2}^L\left(\epsilon_{lLl_2} \tilde{E}_{l_2m_2} - \beta_{lLl_2}\tilde{B}_{l_2m_2}\right),    
\end{equation}
where we have defined $H_{ll_2}^L \equiv {}_+H_{ll_2}^L$ and defined
\begin{equation}
\begin{aligned}
\epsilon_{l L l_2} &\equiv \frac{1+(-1)^{l+L+l_2}}{2}, \\
\beta_{l L l_2} &\equiv \frac{1-(-1)^{l+L+l_2}}{2}.
\end{aligned}
\end{equation}
\Eq{eq:sec5elm} and \Eq{eq:sec5blm} describe the effect of
Faraday rotation on the CMB E-mode and B-mode polarization maps, respectively, which
effectively mixes the multipole moments of the two maps through rotation. This introduces couplings between the E-mode and B-mode maps at different $l$
which otherwise do not exist, given by
\begin{equation}
  \label{app:eq:eb}
  \langle E_{lm}B_{l'm'}^*\rangle_{\rm CMB} = ~\sum_{LM}\alpha_{LM}\xi_{lml'm'}^{LM}f_{lLl'}^{\rm EB},
\end{equation}
with
\begin{equation}
  f_{lLl'}^{\rm EB} = 2\epsilon_{lLl'}[H_{l'l}^L\tilde{C}_l^{\rm EE}-H_{ll'}^L\tilde{C}_l^{\rm BB}].
\end{equation}
The $\langle...\rangle_{\rm CMB}$ denotes that the average is to be taken over CMB
realisations only. One can then define an unbiased quadratic estimator for rotation field as
\begin{equation}
\hat{\alpha}_{LM} = A_L^{\rm EB}\sum_{ll'}\sum_{mm'}\xi_{lml'm'}^{LM}g_{ll'}^{\rm EB}E_{lm}B_{l'm'}^*,
\end{equation}
with the normalization factor to ensure the estimator is unbiased, given by
\begin{equation}
  \left(A_L^{\rm EB}\right)^{-1} = \sum_{ll'}\frac{(2l+1)(2l'+1)}{4\pi}g_{ll'}^{\rm EB}f_{lLl'}^{\rm EB},
\end{equation}
in doing so we have used
\begin{equation}
\sum_{mm'}\xi_{lml'm'}^{LM}\xi_{lml'm'}^{L'M'} = \frac{(2l+1)(2l'+1)}{4\pi}\delta_{LL'}\delta_{MM'}.
\end{equation}
\end{document}